\documentclass[aps,superscriptaddress,twocolumn,preprintnumbers,floatfix,showpacs,prx]{revtex4-1}

\usepackage{amsmath,amssymb}
\usepackage{dcolumn}
\usepackage{braket}
\usepackage{dsfont}
\usepackage{feynmp}
\usepackage{hhline}
\usepackage{graphicx}
\usepackage{mathtools}
\usepackage{xfrac}
\usepackage{paralist}
\usepackage{enumitem}
\usepackage[normalem]{ulem}
\usepackage{color}
\definecolor{darkblue}{rgb}{0.,0.,0.4}
\definecolor{darkred}{rgb}{0.5,0.,0.}

\newcommand{\br}{{\bf r}}
\newcommand{\bR}{{\bf R}}

\newcommand{\beq}{\begin{eqnarray}}
\newcommand{\eeq}{\end{eqnarray}}

\begin{document}

\title{Braiding without Braiding: Teleportation-Based \\ Quantum Information Processing with Majorana Zero Modes}
\author{Sagar Vijay}
\affiliation{Department of Physics, Massachusetts Institute of Technology,
Cambridge, MA 02139, USA}
\affiliation{Kavli Institute for Theoretical Physics, Santa Barbara, CA 93106, USA}

\author{Liang Fu}
\affiliation{Department of Physics, Massachusetts Institute of Technology,
Cambridge, MA 02139, USA}
\begin{abstract}
We present a new measurement-based scheme for performing braiding operations on Majorana zero modes and for detecting their non-Abelian statistics without moving or hybridizing them. In our scheme, the topological qubit encoded in any pair of well-separated Majorana zero modes in a mesoscopic superconductor island is read out from the transmission phase shift in electron teleportation through the island in the Coulomb blockade regime. We propose experimental setups to measure the teleportation phase shift via conductance in an electron interferometer or persistent current in a closed loop.
\end{abstract}
\maketitle


Majorana zero modes are exotic quasiparticle excitations in topological superconductors. Theory predicts that the presence of spatially separated Majorana zero modes gives rise to degenerate superconducting ground states that are indistinguishable by local observables \cite{Kitaev}. Furthermore, braiding  Majoranas is expected to perform a quantized unitary evolution on these ground states, a hallmark of their non-Abelian statistics \cite{moore-read, read-green, ivanov}. Due to these remarkable properties, Majorana zero modes have been proposed as topological qubits for robust quantum information processing that is (ideally) error-free at zero temperature \cite{Beenakker, Alicea, Nayak-rmp}.

Following theoretical proposals, over the last few years transport and scanning tunneling microscopy experiments  have reported the observation of zero-bias conductance peak as a signature of Majorana zero modes in various material platforms including nanowires \cite{Delft}, atomic chains \cite{Princeton} and topological insulators \cite{Vortex_Tunneling}, proximitized by $s$-wave superconductors. These results suggest the existence of Majorana zero modes, and encourage research towards demonstrating their topological properties. Among these, non-Abelian statistics is widely regarded as the ``holy grail" for topological phases of matter and for topological quantum computation.

Theoretical proposals for detecting the non-Abelian statistics of Majoranas have mostly relied on braiding, i.e. moving Majoranas around each other via a sequence of operations. For example, by changing the phase of Josephson junctions, Majorana zero modes localized in Josephson vortices can be braided in an array of superconducting islands on a topological insulator \cite{Fu_Kane}. By tuning the gate voltage, Majoranas in proximitized nanowires can be braided in a T-junction \cite{T_Junction, Milestone_Majorana}. Detecting non-Abelian statistics further requires measuring the state of Majoranas before and after braiding. Both the motion and measurement of Majoranas are yet to be experimentally achieved. Furthermore, physically moving Majoranas in nanowire networks suffers from dangerous thermal errors that are very difficult to correct \cite{DiVincenzo}.  These errors may be avoidable in other proposals that selectively tune couplings between Majoranas to implement braiding transformations \cite{Sau, Akhmerov, Oppen}.

In this work, we introduce a new scheme for (i) detecting the non-Abelian statistics of Majorana zero modes and (ii) implementing braiding operations, without any physical braiding, which is entirely based on projective measurement as opposed to unitary evolution. In our scheme, a topological qubit encoded in any pair of well-separated Majoranas is read out from the transmission phase shift in electron teleportation through the topological superconductor that hosts these Majoranas \cite{Teleportation}. Electron teleportation is a remarkable mesoscopic transport phenomenon enabled by the fractional nature of Majorana zero modes and the charging energy of the superconductor. Here we use electron teleportation to directly measure and manipulate Majorana qubits without moving, hybridizing or destroying Majorana zero modes.

In our scheme for ``braiding without braiding",
 the unitary transformation that would be generated by physically exchanging a pair of Majoranas is realized by performing a sequence of projective measurements of Majorana bilinear operators. The theoretical basis for using projective measurements to implement quantum gates was provided in Ref. \cite{Nielsen, Nielsen_Chuang}.  Within the abstract setting of non-Abelian topological order, replacing anyon braiding by topological charge measurements was proposed by Bonderson, Freedman and Nayak \cite{Bonderson}. On the other hand, electron teleportation provides an ideal way of measuring Majorana qubits in mesoscopic topological systems, where the charging energy required for teleportation comes from the long-range Coulomb interaction. As a result, the physics of teleportation lies beyond the theory of topological order in the thermodynamic limit. By combining teleportation-based measurement and measurement-based braiding, our work unveils a novel approach to quantum information processing with well-separated, stationary Majorana zero modes.

Our work is especially timely in view of a recent groundbreaking experiment on epitaxially grown InAs/Al superconducting nanowires \cite{Copenhagen_2}, which are theoretically predicted to host Majorana end modes under an external magnetic field \cite{Nanowire_Proposal, oreg, palee}. Due to charging effects in the Coulomb blockade regime, transport through the nanowire at zero magnetic field is dominated by Cooper pair tunneling, leading to zero-bias conductance oscillations with the gate voltage that are charge-$2e$ periodic.  However, above a critical field and in the presence of a superconducting gap, the conductance oscillations become charge-$e$ periodic. The observed charge-$e$ transport in a superconducting state supports the theoretically predicted scenario of electron teleportation via Majorana modes \cite{Teleportation, Egger-PRL, Glazman_1}. Another distinctive feature of teleportation is that single-electron transport through the superconducting island is phase coherent \cite{Teleportation}. This important property forms the basis for topological qubit readout in this work. To detect the phase coherence requires an electron  interferometer, which is currently under experimental pursuit \cite{Marcus-private}. Given these exciting developments, we believe teleportation-based braiding without braiding is a practical scheme for detecting the non-Abelian statistics of Majorana zero modes, and offers a promising prospect for quantum information processing.

Our teleportation-based scheme for implementing projective measurements and performing ``braiding without braiding" on stationary and spatially-separated Majorana zero modes has significant advantages over other schemes based on physically moving Majoranas to implement logical gates or to perform qubit readout. For braiding to be feasible, Majoranas must be moved sufficiently slowly to obey an adiabaticity condition \cite{Milestone_Majorana}, which is especially stringent in disordered nanowires without a hard spectral gap \cite{Sondhi, Rahul}. Qubit readout and gate operation in our proposal are not limited by this constraint. Moreover, in the process of moving Majoranas, dangerous thermal errors on the topological qubit may be accumulated, which are extremely difficult (if not impossible) to de-code and correct \cite{DiVincenzo_2}.  Finally, teleportation-based measurement of Majorana qubits has advantages over proposed readout schemes based on charge sensing \cite{Milestone_Majorana} which can only be performed on pairs of Majorana zero modes that are spatially adjacent.

Our paper is organized as follows.  We begin by reviewing the phenomenon of phase-coherent electron teleportation through Majorana zero modes.  We then describe two teleportation-based setups -- the Majorana interferometer and the Majorana SQUID -- for measuring a topological qubit encoded in a pair of well-separated Majorana zero modes, and for detecting their non-Abelian statistics.  Next, we present a general protocol for implementing braiding transformations on Majorana zero modes exclusively through projective measurements.  Finally, we provide a concrete experimental realization of our proposal using proximitized nanowires. Our general scheme of teleportation-based braiding without braiding is applicable to any Majorana platform, provided that the topological superconductor hosting the Majoranas has a finite charging energy.

\section{Conceptual Basis}
In this section we lay out the theoretical basis of teleportation-based measurement of a topological qubit encoded in a pair of spatially separated Majorana zero modes. We first elaborate on the transmission phase shift in electron teleportation via a pair of Majoranas and its dependence on the state of the topological qubit, as pointed out in Ref. \cite{Teleportation}. Next, we propose two ways of detecting this phase shift, or equivalently reading out the topological qubit, by measuring the conductance in an electron interferometer or the persistent current in a closed loop. We then explicitly show the change of the teleportation phase shift in the process of physically exchanging two Majorana zero modes. The difference in the phase shift---a physical observable measured by interferometry---before and after the braiding directly proves the system has evolved into a new state, thus demonstrating the
non-Abelian statistics of Majorana zero modes.

\subsection{Teleportation-Based Measurement of Topological Qubit}

Let us consider a mesocopic topological superconductor island hosting a number of well-separated Majorana zero modes that have negligible wavefunction hybridization. Each Majorana zero mode of interest is tunnel coupled to a normal metal lead, and the tunnel couplings can be turned on and off by gates. The superconducting island is capacitively coupled to a nearby gate. We assume that the charging energy $E_c$ is smaller than the superconducting gap $\Delta$, but larger than the tunnel coupling to the leads, as defined by Eq.(\ref{Gamma}) below.

In the absence of a tunnel coupling to leads,  the ground state energy of the island depends on the total number of electrons $N$ through the charging energy:
\beq
E(N) = E_c (N-n_g)^2,
\eeq
where the offset charge $n_g$ is continuously tunable by the gate voltage.
Due to the presence of Majorana zero modes, the superconducting island can accommodate an even and an odd number of electrons on equal ground without paying the energy cost of the superconducting gap. Thus $N$ takes both even and odd integer values. Throughout this work, we assume that the island is in the Coulomb blockade regime away from the charge degeneracy point, so that the total charge of the island is fixed, denoted by $N=N_0$. Under this condition, the island has $2^{M/2-1}$ degenerate ground states, where $M$ (an even integer) is the number of Majorana zero modes present. These degenerate ground states form a topologically protected Hilbert space, which we use to encode quantum information.
By detuning the island far away from charge degeneracy, the topological qubits are protected against quasiparticle poisoning from outside the island at low temperature.

A complete basis for this $2^{M-1}$-dimensional Hilbert space is given by the common eigenstates of a set of nonoverlapping Majorana bilinear operators, e.g., $( i \gamma_1 \gamma_2, i\gamma_3 \gamma_4, ..., i \gamma_{M-1} \gamma_M)$. A Majorana bilinear operator $i \gamma_a \gamma_b$ has two eigenstates $|\pm\rangle_{ab}$, defined by
\beq
i \gamma_a \gamma_b | \pm \rangle_{ab} = \pm | \pm \rangle_{ab}.
\eeq
Thus, measuring the topological qubit in this basis amounts to measuring the eigenvalue of $i \gamma_a \gamma_b$.
Note that any way of partitioning Majoranas into pairs defines a corresponding basis for the topological qubit, and  different bases are related by unitary transformations known as $F$-symbols, which are determined by the fusion rule of Majoranas. It is thus highly desirable to measure the eigenvalue of {\it any} Majorana bilinear operator, so that the topological qubit can be measured equally well in any basis.

\begin{figure}[t]
$\begin{array}{ccc}
\includegraphics[trim = 0 -11 0 0, clip = true, width=0.21\textwidth, angle = 0.]{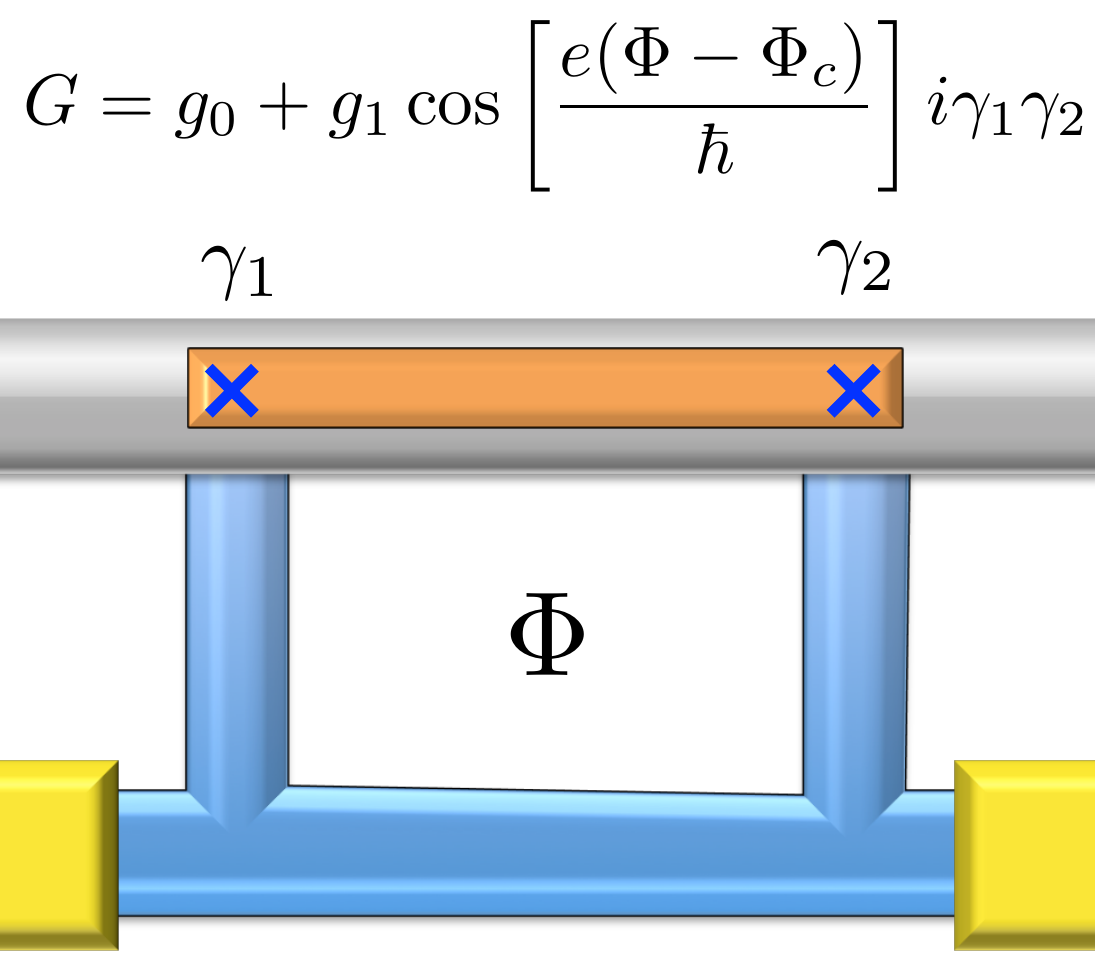} & &
\,\,\,\,\includegraphics[trim = 0 0 0 0, clip = true, width=0.24\textwidth, angle = 0.]{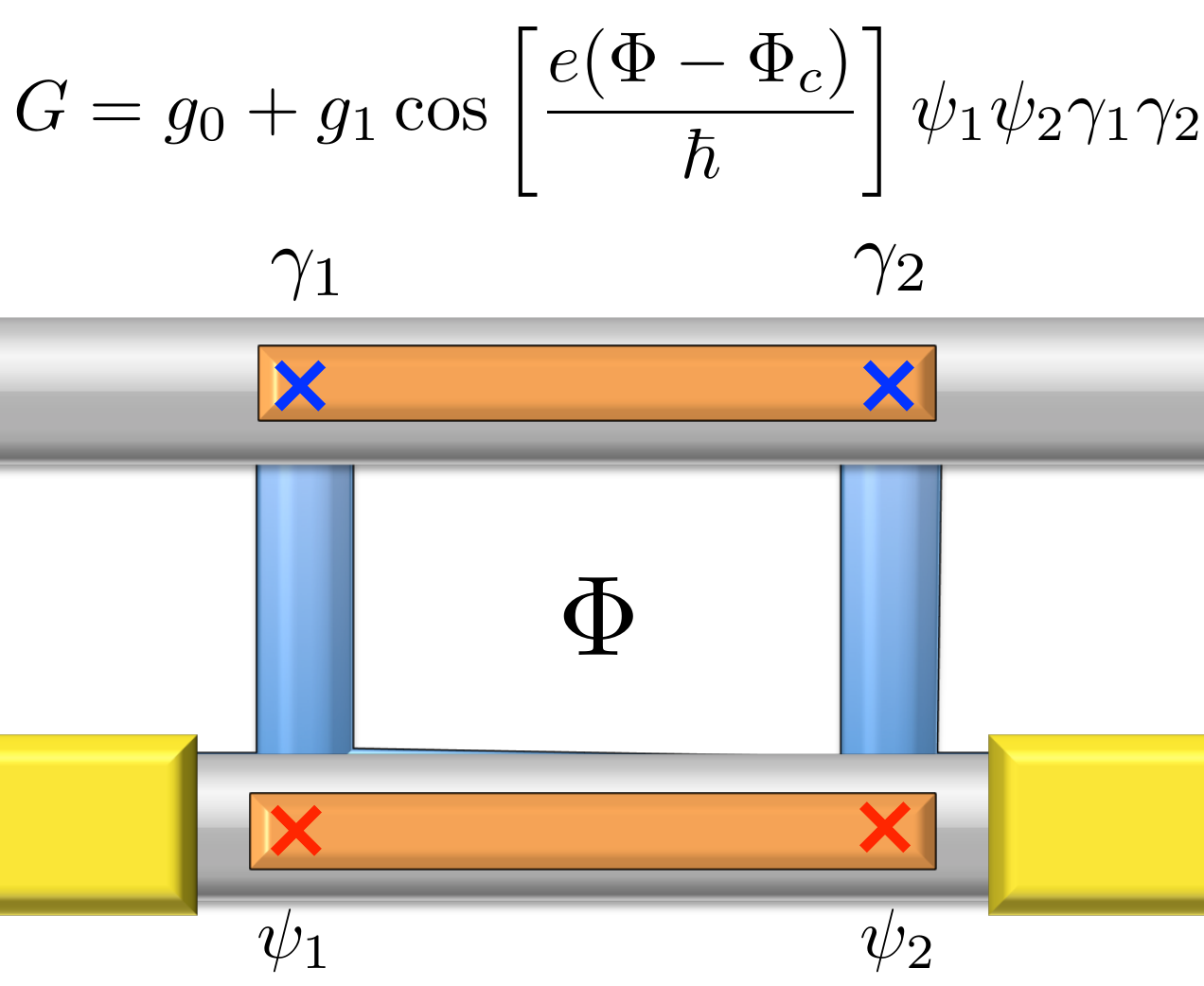}\\\\
\text{(a)} & & \,\,\,\,\text{(b)}
\end{array}$
\caption{{\bf Majorana Interferometer --} Two electron interferometry setups to measure the topological qubit formed by Majorana zero modes $\gamma_{1}$ and $\gamma_{2}$.  In both interferometers, one path goes through the topological qubit while the other path goes through (a) a normal metal with sufficiently long phase coherence length (blue) and (b) a second Majorana island initialized in a definite parity state $i\psi_{1}\psi_{2} = \pm 1$, which is used as a reference.   }
  \label{fig:Interferometry}
\end{figure}

We now describe a teleportation-based protocol to measure the eigenvalue of any Majorana bilinear $i \gamma_a \gamma_b$ by coupling the Majorana island to lead $a$ and to lead $b$. The bare tunneling Hamiltonian is given by
\beq
H^0_T= \sum_{j=a,b} t_j c_j^\dagger(0) f(\br_j) + \mathrm{h.c.} \label{tunnel}
\eeq
where $c_j(0)$ is the electron operator at the end of lead $j$, and $f(\br_j)$ is the electron operator in the island at the tunneling location $\br_j$, where the Majorana zero mode $\gamma_j$ is located. Next, we expand $f(\br_j)$ in terms of quasiparticle operators in the superconducting island:
\beq
f^\dagger(\br_j) = \xi^*_j(\br) e^{i\theta/2} \gamma_j + ... \label{f}
\eeq
Here $\xi_j(\br)$ is the wavefunction associated with the Majorana mode operator $\gamma_j$, defined by
\beq\label{eq:gamma_def}
\gamma_j = \int d \br \; \left[\xi_j(\br) e^{-i\theta/2} f^\dagger(\br) + \xi^*_j(\br) e^{i\theta/2}  f(\br)\right]. \label{gamma}
\eeq
Here, $\theta$ is the phase operator of the superconductor, which is conjugate to the electron number operator $N$ and satisfies the commutation relation $[\theta, N] =2 i$.
In the operator expansion (\ref{f}) we have neglected all quasiparticles above the superconducting gap which are irrelevant to the low energy physics of our interest, as well as Majorana zero modes at other locations whose amplitudes at $\br_j$ are exponentially small.
Thus, as shown by (\ref{f}), in the low-energy Hilbert space the electron creation operator $f^\dagger(\br_j)$ is represented as a product of the Majorana mode operator $\gamma_j$ and the charge-raising operator $e^{i\theta/2}$ which increases the charge of the island by $1e$. Physically speaking, Eq. (\ref{f}) describes the charge-statistics separation of an electron after entering a topological superconductor: the charge of the electron is spread out over the entire superconductor, while its Fermi statistics is retained by a localized Majorana fermion that is charge neutral.

Substituting (\ref{f}) into the bare Hamiltonian (\ref{tunnel}) yields an effective tunneling Hamiltonian
\begin{align}
H_T &= \sum_{j=a,b} \lambda_j c^\dagger_j (0) \gamma_j e^{-i\theta/2}+ \mathrm{h.c.}, \textrm{ with }\lambda_j =t_j \xi_j(\br_j) \nonumber
\end{align}
We define the tunnel coupling $\Gamma$ as
\beq
\Gamma = 
\sum_{j=a,b} 2\pi\rho {\lambda}_j^2, \label{Gamma}
\eeq
where  $\rho$ is the density of states in the leads.  Assuming $\Gamma \ll E_\pm$ where $E_\pm \equiv E(N_0\pm 1) - E(N_0)$ is the energy difference between the charge states $N=N_0$ and $N=N_0\pm 1$, transmission through the island is dominated by a second-order process, where a single electron tunnels into the island from one lead and a single electron exits from the island to another lead. Therefore, from second-order perturbation in $H_T$, we obtain an effective coupling between a Majorana island in the off-resonance Coulomb blockade regime and the leads
\begin{widetext}
\beq
H_{ab} &=&  - \lambda_a^* \lambda_b  c_b^\dagger(0) c_a(0) \left[ \frac{ \langle N_0|  \gamma_b e^{-i\theta/2} | N_0+1 \rangle
\langle N_0+1  |  \gamma_a e^{i\theta/2}|  N_0 \rangle} {E(N_0+1) - E(N_0)}
+ \frac{ \langle N_0|  \gamma_a e^{i\theta/2} | N_0-1 \rangle
\langle N_0-1  |   \gamma_b e^{-i\theta/2} |  N_0 \rangle} {E(N_0-1) - E(N_0)} \right] + \mathrm{h.c.} \nonumber \\
&=&   \gamma_a \gamma_b \left[ T_{ab} c_b^\dagger(0) c_a (0)- T_{ab}^* c_a^\dagger (0) c_b(0) \right],   \label{Aij}
\eeq
\end{widetext}
where $T_{ab}\equiv \lambda_a^* \lambda_b \left(\frac{1}{E_+}+ \frac{1}{E_-} \right)$ is the effective single electron tunneling between lead $a$ and $b$, mediated by a pair of Majorana zero modes $\gamma_a, \gamma_b$.
Due to this entanglement of Majorana degrees of freedom with electron tunneling between two leads, $H_{ab}$ enables a direct projective measurement of the Majorana bilinear $i\gamma_a \gamma_b$, even when $\gamma_a$ and $\gamma_b$ are far apart in the superconductor island, as we show below.

\begin{figure}[t]
$\begin{array}{ccc}
\includegraphics[trim = 0 -11 0 0, clip = true, width=0.2\textwidth, angle = 0.]{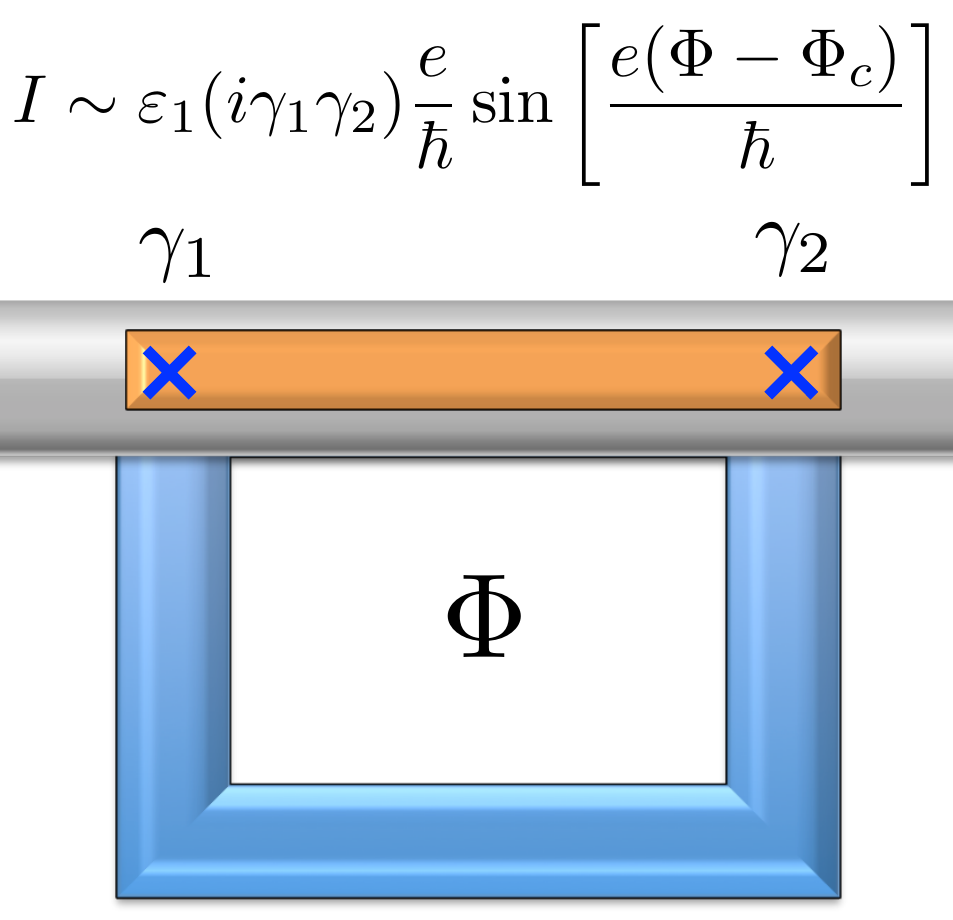} & &
\,\,\,\,\includegraphics[trim = 0 0 0 0, clip = true, width=0.23\textwidth, angle = 0.]{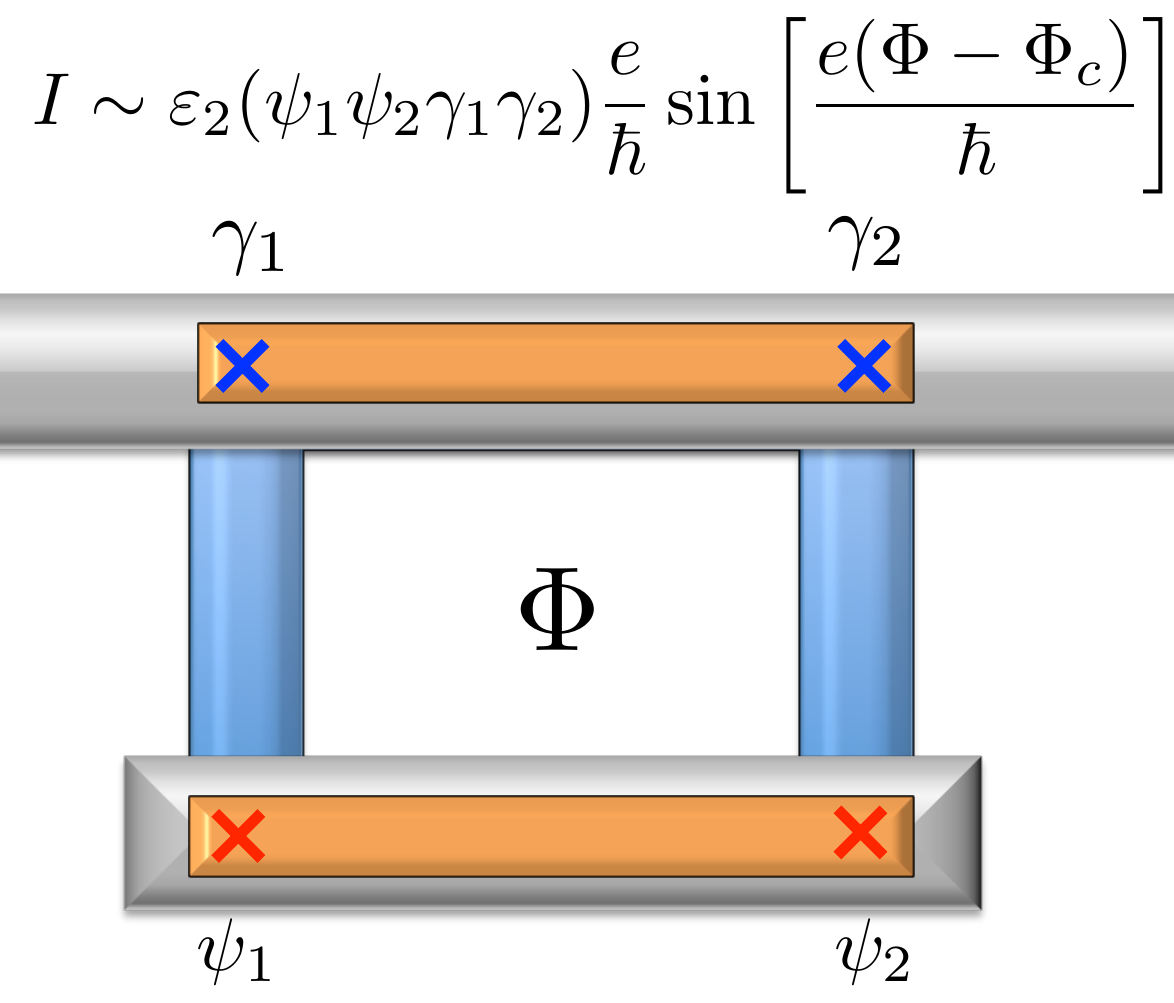}\\\\
\text{(a)} & & \,\,\,\,\text{(b)}\\\\
\end{array}$
\caption{{\bf Majorana SQUID --} When the two Majorana zero modes $\gamma_{1}$ and $\gamma_{2}$ are connected by a bridge outside the island to form a closed loop, with the bridge being (a) a normal metal with sufficiently long phase coherence length or (b) a reference Majorana island in a definite parity state $i\psi_{1}\psi_{2} = \pm 1$, the topological qubit defined by $i\gamma_1 \gamma_2=\pm 1$ may be read out by measuring the persistent current $I$ in the ground state, which is a $h/e$-periodic  function of the applied flux $\Phi$. }
  \label{fig:SQUID}
\end{figure}

Let us first consider the case that the Majorana island is initialized to be an eigenstate of $i\gamma_a \gamma_b$, either $|+\rangle_{ab}$ or $|-\rangle_{ab}$. It follows from (\ref{Aij}) that the single electron tunneling amplitude from lead $a$ to $b$, which is mediated by $\gamma_a$ and $\gamma_b$, is equal to $-i T_{ab}$ for the Majorana qubit state  $|+\rangle_{ab}$, and $+i T_{ab}$ for the state $|-\rangle_{ab}$. Therefore, the two Majorana qubit states $|\pm\rangle_{ab}$ are distinguishable by the $\pi$ difference in the transmission phase shift in electron teleportation via a pair of Majorana zero modes \cite{Teleportation}.


To measure the teleportation phase shift requires quantum interference. We now propose two phase measurement schemes for Majorana qubit readout. The first scheme is based on a conductance measurement in a two-path electron interferometer, with one path going through the Majorana island and the other path serving as a reference. The reference path may be a normal metal with a sufficiently long phase coherence length \cite{Teleportation}, or a second Majorana island in a definite parity state \cite{Xu-Fu, Egger}, as shown in Fig. \ref{fig:Interferometry}. The total conductance $G$ then contains a term proportional to $ (i \gamma_a \gamma_b)$ due to the interference between the two paths, i.e., $G(\Phi) = g_0 + i g (\Phi) \gamma_a \gamma_b$, where
$g$ depends periodically on the external magnetic flux $\Phi$ enclosed by the two interfering paths, with $h/e$-periodicity. Since the conductance takes different values for the qubit state $|\pm \rangle_{ab}$, the conductance measurement in such Majorana interferometer provides a projective measurement of the topological qubit in the basis $|\pm \rangle_{ab}$.

The second scheme for qubit readout is based on measuring the persistent current in a closed loop. This loop can be made by connecting  Majorana zero modes on the island to the ends of a normal metal bridge (see Fig. \ref{fig:SQUID}a), or to a reference  Majorana island in a definite qubit state (see Fig. \ref{fig:SQUID}b). Due to the phase coherence of  electron motion around the loop, the energy of the closed system depends periodically on the external magnetic flux $\Phi$ through the loop with $h/e$ periodicity,
\beq
E = E_0 + i \varepsilon \gamma_a \gamma_b  \cos \left[  \frac{e(\Phi - \Phi_c)}{\hbar} \right], \label{ephi}
\eeq
where $\Phi_c$ and $\varepsilon$ depend on details of the setup such as tunnel couplings between the island and the normal metal bridge.
Eq.(\ref{ephi}) implies the presence of a persistent circulating current in the loop
\beq
I=\frac{\partial E}{\partial \Phi} =  (i \gamma_a \gamma_b) \frac{e \varepsilon}{\hbar} \sin \left[\frac{e (\Phi-\Phi_c)}{\hbar}\right].
\eeq
This circulating current flows in opposite directions for the two Majorana qubit states $|\pm\rangle_{ab}$. Thus the Majorana qubit is faithfully transferred to the state of the persistent current, which can then be read out by inductive coupling the system to a SQUID loop.

We now estimate the magnitude of the persistent current in a Majorana SQUID by treating the transmission through a Majorana island as single electron hopping across a weak link, as described by the effective Hamiltonian (\ref{Aij}).  Details of our calculation are presented in the Supplemental Material \cite{SM}.  When the Majorana SQUID is formed by a single island connected to a normal metal bridge, we find that the magnitude of the persistent current at zero temperature is given by
\begin{align}
I_{0} \sim \frac{ e\Gamma}{\hbar} \delta\left(\frac{1}{E_+} + \frac{1}{E_-}\right)
\end{align}
as explicitly calculated in the Supplemental Material \cite{SM}. Here, $\Gamma$ is the tunnel coupling between the island and the normal metal defined in (\ref{Gamma}), and $\delta$ is the single-particle level spacing in the metal, which is inversely proportional to the length of the bridge.
An order-of-magnitude estimate based on experimental parameters in Ref. \cite{Copenhagen_2, Copenhagen,  Glazman_1} yields $I_{0} \sim 10$ nA.

When the Majorana SQUID consists of two islands connected by two normal metal bridges, we determine the persistent current by modeling the bridges as mediating a direct electron tunneling between the Majorana islands.   In this case, we consider the Hamiltonian $H = H_{T} + H_{c}$ for the full system, where
\begin{align}\label{eq:H_c}
H_{c} = \sum_{i=1,2}E_{c}^{(i)}(N_{i} - n_{g}^{(i)})^{2}
\end{align}
describes the charging energy for each of the Majorana islands ($i = 1$, $2$).  Here, $E_{c}^{(i)}$, $N_{i}$ and $n_{g}^{(i)}$ are the charging energies, total charge, and gate charges, respectively, for island $i$.  For simplicity, we let $E_{c}^{(1)} = E_{c}^{(2)} = E_{c}$ for the remainder of our calculation. Furthermore, as shown in Sec. IA, electron tunneling between the two Majorana islands is described at low energies by an effective Hamiltonian in terms of the Majorana operators, as given by
 \begin{align}\label{eq:H_J}
H_{t} =    &\,\,\,it_{1} \psi_{1}\gamma_{1} \,e^{i (\theta_1 - \theta_2)/2}  + \mathrm{h.c.}\nonumber\\
&+ i t_{2} e^{i e\Phi/\hbar} \psi_{2}\gamma_{2}\, e^{i (\theta_2 -\theta_1)/2} + \mathrm{h.c.}
\end{align}
with $\theta_{1,2}$ the superconducting phases on each Majorana island and $\Phi$, the applied flux through the ring.

In the presence of a large charging energy $E_{c} \gg t_{1,2}$, the effective Hamiltonian for the system is, to lowest order in perturbation theory, given by
\begin{align}
H_{\mathrm{eff}} = \frac{2 |t_{1}t_{2}|}{E_{+} + E_{-}}\cos\left[\frac{e(\Phi-\Phi_c)}{\hbar}\right]\psi_{1}\psi_{2}\gamma_{1}\gamma_{2}, \label{Hct}
\end{align}
where the constant $\Phi_c$ provides an overall shift and is present when $t_{1,2}$ are complex.
The magnitude of the persistent current is then given by
\begin{align}
I_{0} \sim \frac{e}{\hbar}\left(\frac{|t_{1}t_{2}|}{E_{+} + E_{-}} \right)
\end{align}
In the Supplemental Material \cite{SM}, we also model the persistent current in a Majorana SQUID with two Majorana islands as single electron hopping in a ring with two weak links and determine the magnitude of the persistent current \cite{SM}. 

\subsection{Detecting non-Abelian Braiding Statistics from Teleportation Phase Shifts}
In this section we explicitly demonstrate the change of teleportation phase shift due to braiding Majorana zero modes in a two-dimensional topological superconductor. Here, Majorana braiding is realized by adiabatically exchanging two identical vortices, which host Majorana zero modes in their cores \cite{read-green}. Since the teleportation phase shift is a physical observable that can be measured by interferometry, its change before and after braiding implies a change in the quantum state of the system, thus providing direct proof for non-Abelian statistics.

Before proceeding, we first clarify what we mean by non-Abelian statistics of Majorana-carrying vortices in a superconductor. We assume that the superconductor is well described by a Bogoliubov-de Gennes (BdG) Hamiltonian with a pairing potential $\Delta(\br)=|\Delta(\br)| e^{i \theta(\br)}$ that is a complex function of position. We assume that apart from the overall phase $\theta$, the pairing potential configuration $\Delta(\br)$ is non-dynamical and externally set up. On the other hand, we take the overall superconducting phase $\theta$ as a quantum mechanical variable, which is conjugate to the total number of electrons $N$. Throughout this work, we take $N$ to be fixed due to the large charging energy, so that $\theta$ is fluctuating.

In this setting, Majorana zero modes are not deconfined anyons but a type of ``twist defect'' \cite{Teo} associated with vortices, the point singularities in $\Delta(\br)$. A vortex centered at $\bR$ corresponds to a $\pm 2\pi$ winding of the phase $\theta(\br)$ around $\bR$. As we adiabatically exchange two vortices, $\Delta(\br)$ varies slowly. To define non-Abelian statistics, it is required that the full function $\Delta(\br)$ returns to its original configuration after the vortex exchange. 
The evolution of the system into a new quantum state after this process is a defining feature of the non-Abelian statistics of Majorana zero modes bound to vortex cores.

\begin{figure}
$\begin{array}{c}
\includegraphics[trim = 0 0 0 0, clip = true, width=0.34\textwidth, angle = 0.]{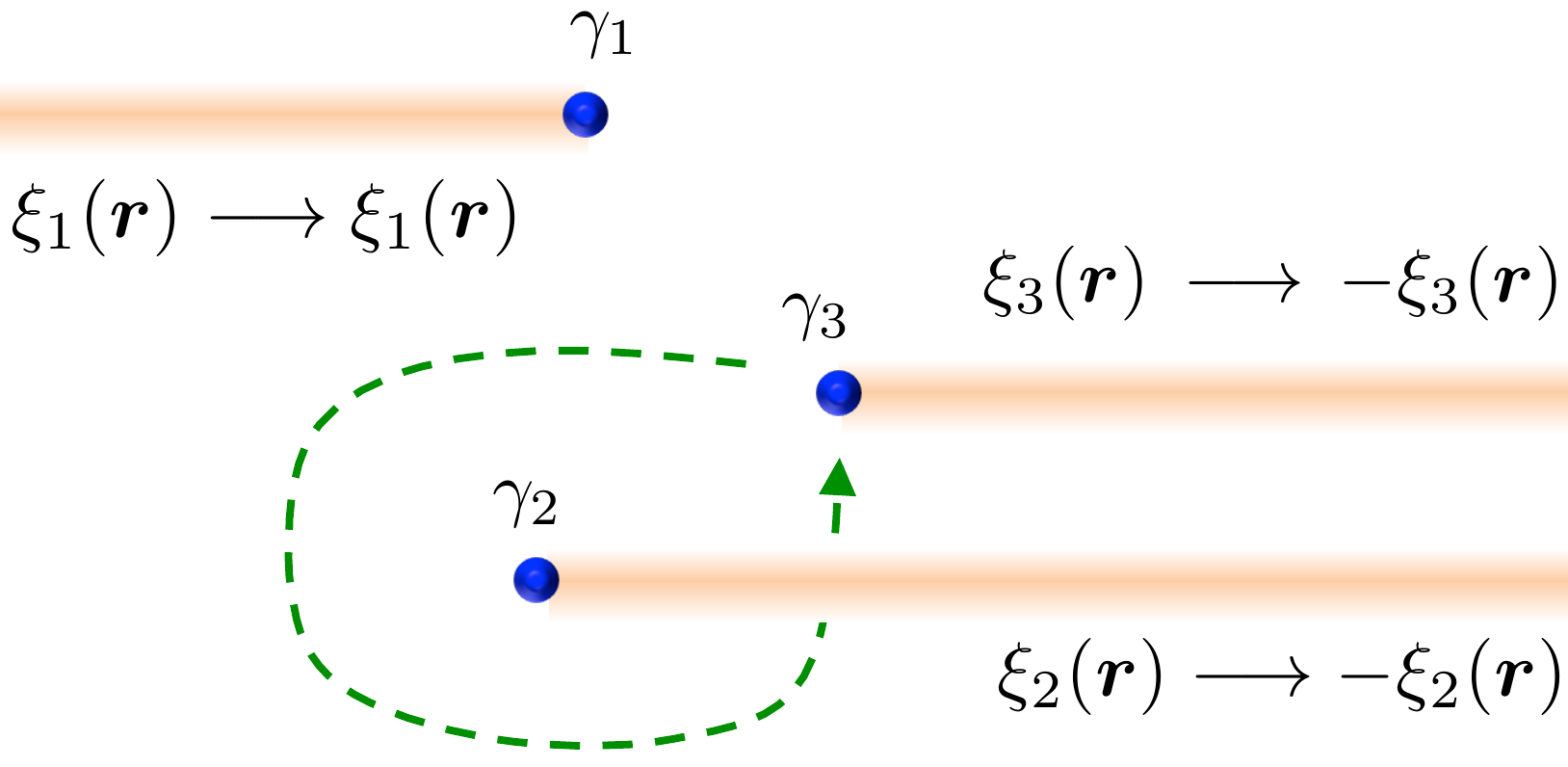}\\
\text{(a)}\\\\\\
\includegraphics[trim = 0 0 0 0, clip = true, width=0.34\textwidth, angle = 0.]{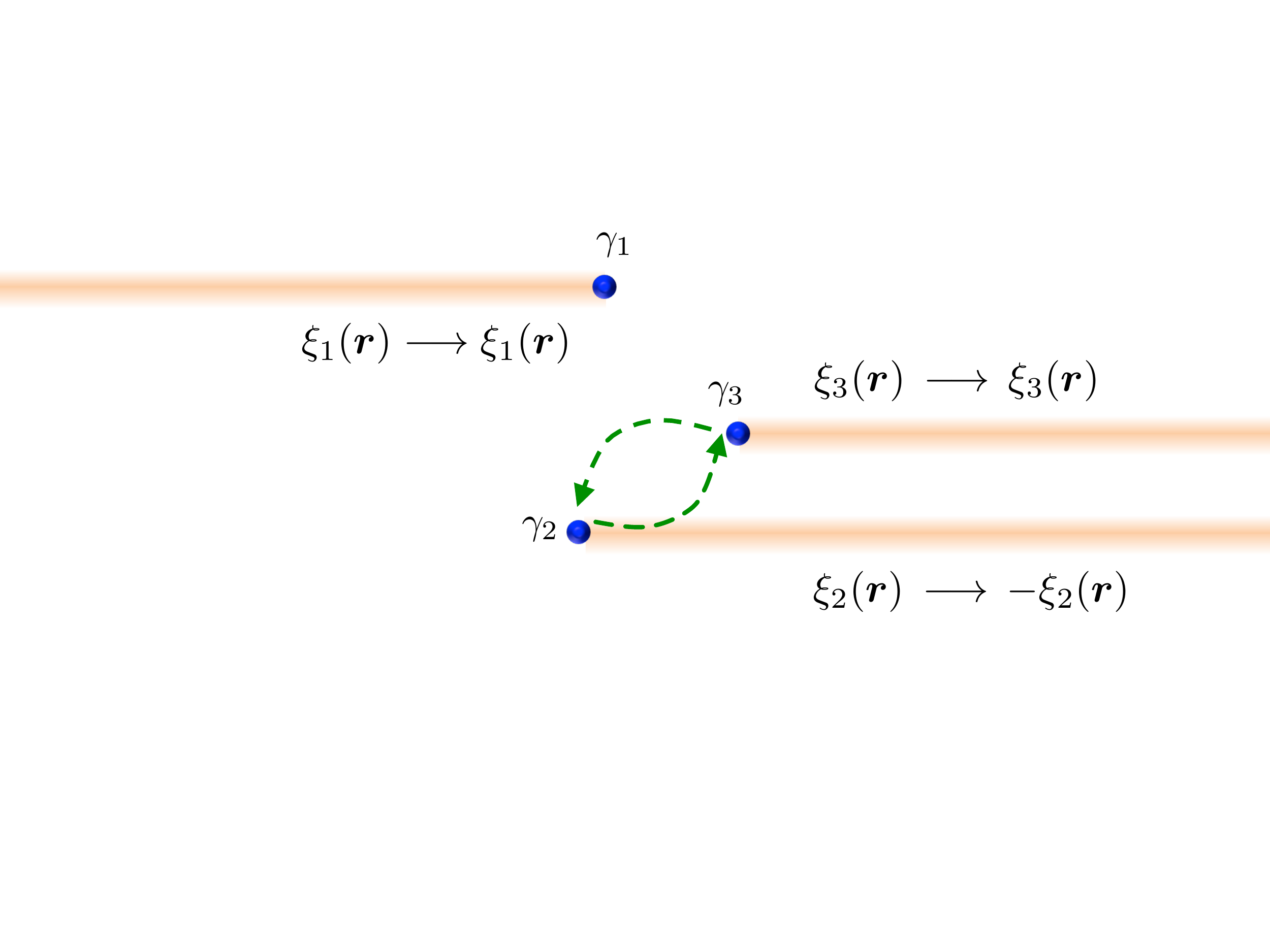}\\
\text{(b)}
\end{array}$
\caption{{\bf Teleportation Phase-Shift --} Braiding (a) or exchanging (b) Majorana zero modes induces a transformation on the wavefunctions as indicated.  The overall sign of all of the wavefunctions is ambiguous, and only gauge-invariant quantities, given by products of even numbers of Majorana wavefunctions, appear in physical observables.  The shaded lines shown above are physical regions where the superconducting phase rapidly changes by $2\pi$.  The change in sign of physical observables, due to the ``branch cuts" sweeping through the Majorana zero modes \cite{ivanov}, provides a signature of their non-Abelian statistics that may be detected via teleportation.}
  \label{fig:Braiding_1}
\end{figure}

As a warm-up, consider two well-separated vortices centered at $\bR_1$ and $\bR_2$, and denote the corresponding Majorana zero modes localized in the vortex cores by $\gamma_1$ and $\gamma_2$.  We connect $\gamma_1$ and $\gamma_2$ by a normal metal bridge to form an interferometer as discussed in Sec. IA, and consider how  the teleportation phase shift evolves as a third vortex moves around the vortex at $\bR_2$ in a full circle (see Fig. \ref{fig:Braiding_1}a).

For any given vortex configuration, the wavefunction associated with any Majorana zero mode $\xi_j(\br)$, obtained by solving the BdG Hamiltonian, is defined up to an overall choice of sign. Since the Majorana zero mode operator $\gamma_{j}$ is defined from $\xi_{j}(\br)$ via Eq. (\ref{gamma}), $\gamma_j$ is {\it not gauge-invariant} as emphasized in Ref. \cite{Halperin}.
Nonetheless, this choice of sign for the wavefunctions $\xi_{1,2}(\br)$ does {\it not} affect any physical observables, which necessarily correspond to gauge-invariant operators such as $\xi_{j} (\br)\gamma_j$.

For convenience, we now choose the signs of these wavefunctions such that $\xi_{1,2}(\br)$ vary continuously with the moving position of the third vortex. The eigenvalue of the Majorana bilinear operator $i\gamma_1 \gamma_2$, 
taking two possible values $\pm 1$, stays constant during the braiding process, 
as the fermion parity of the system is conserved.
As a result, the teleportation amplitude, whose expression (\ref{Aij}) contains the product $\xi_{1}(\bR_1)\xi_2(\bR_2) \gamma_1 \gamma_2$, also varies continuously.  As shown by Ivanov \cite{ivanov}, after the third vortex returns to its original position and the original vortex configuration is restored, the wavefunction $\xi_1(\br)$ comes back to itself while $\xi_2(\br)$ and $\xi_3(\br)$ change sign, as shown in Fig. \ref{fig:Braiding_1}a. Consequently, the phase shift in electron teleportation via Majorana zero modes $\gamma_1$ and $\gamma_2$ changes by $\pi$ before and after the third vortex circles around $\gamma_2$. This quantized change of a physical observable signals a change in the quantum state of the system induced by braiding.

The teleportation phase can also detect the change in the state of the system when two vortices are \emph{exchanged}, as shown in Fig. \ref{fig:Braiding_1}b. We assume that the local configurations of the pairing potential near the vortex centers $\bR_2$ and $\bR_3$ are identical, so that the wavefunctions of the Majorana zero modes $\gamma_2$ and $\gamma_3$ are essentially related by translation, i.e. $\xi_2(\br-\bR_2)=\xi_3(\br-\bR_3)$. After exchanging vortices 2 and 3 in the manner shown in Fig. \ref{fig:Braiding_1}b and restoring the original vortex configuration, the Majorana wavefunctions transform as
$\xi_{2}(\br) \longrightarrow - \xi_{2}(\br)$, and
$\xi_{3}(\br) \longrightarrow \xi_{3}(\br)$. To demonstrate the braiding-induced change in the quantum state of the system, we connect $\gamma_2$ and a reference Majorana zero mode $\gamma_1$ by a normal metal bridge to form an interferometer, and monitor the evolution of the teleportation phase shift in the process of exchanging $\gamma_2$ and $\gamma_3$, while keeping one end of the bridge attached to the moving Majorana $\gamma_2$. The initial teleportation amplitude from $\bR_1$ to $\bR_2$ is given by
the product $\xi_1(\bR_1) \xi_2(\bR_2) \gamma_1 \gamma_2$. After braiding, this interferometer measures the teleportation amplitude from $\bR_1$ to $\bR_3$, given by $\xi_1(\bR_1) \xi_3(\bR_3) \gamma_1 \gamma_2$. This result should be compared with the teleportation amplitude from $\bR_1$ to $\bR_3$ before braiding, given by $\xi_1(\bR_1) \xi_3(\bR_3) \gamma_1 \gamma_3$ and measurable by an interferometer containing $\gamma_1$ and $\gamma_3$. This comparison shows that braiding $\gamma_2$ and $\gamma_3$ has the effect of the transformation $\gamma_3  \rightarrow \gamma_2$.
Repeating the same analysis for the teleportation amplitude from $\bR_1$ to $\bR_2$ shows that same braiding process also has the effect of the transformation $\gamma_2 \rightarrow -\gamma_3$. Our analysis based on electron teleportation thus reproduces the ``Ivanov rule'' for Majorana braiding  \cite{ivanov}
\beq
\gamma_2 \rightarrow -\gamma_3, \gamma_3 \rightarrow \gamma_2. \label{rule}
\eeq

It is worth noting that the above braiding transformation (\ref{rule}) {\it per se} is non-gauge-invariant, as it is expressed in terms of Majorana operators that suffer from a$Z_2$ sign ambiguity. Only after the sign convention for the zero mode wavefunction $\xi_{2,3}$ is specified, as we did previously by choosing $\xi_2(\br-\bR_2)=\xi_3(\br-\bR_3)$, do the Majorana operators $\gamma_{1,2}$ become well-defined, so that the braiding transformation (\ref{rule}) becomes meaningful.

Our analysis, as presented, demonstrates that the braiding-induced change in the teleportation phase shift is a physical observable described by a gauge invariant operator involving $\xi^*_{a}\xi_b \gamma_a \gamma_b$ and $\xi_a \xi_b^*\gamma_a \gamma_b$. Thus the change is teleportation phase shift is a {\it direct} and {\it measurable} consequence of the non-Abelian statistics of Majorana zero modes.



\section{Measurement-Based Braiding}
\begin{figure}
\includegraphics[trim = 0 0 0 0, clip = true, width=0.34\textwidth, angle = 0.]{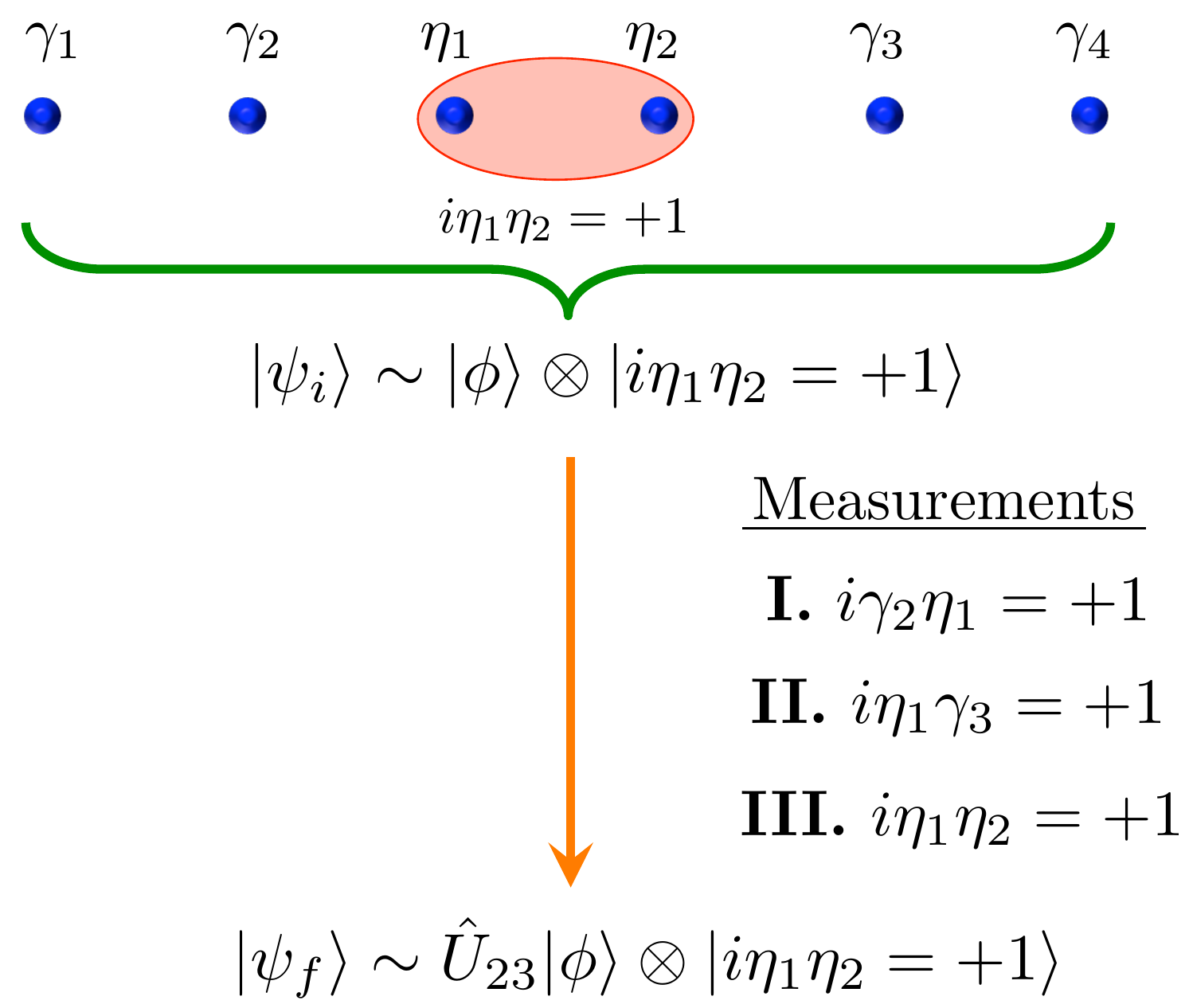}
\caption{{\bf Measurement-Based Braiding --} Schematic depiction of the initial state $\ket{\psi_{i}}$, with Majorana zero modes $\eta_{1}$ and $\eta_{2}$ initialized in the state $i\eta_{1}\eta_{2} = +1$.  Performing the indicated sequence of measurements is equivalent to braiding Majorana fermions $\gamma_{2}$ and $\gamma_{3}$, up to a normalization factor.}
  \label{fig:Schematic}
\end{figure}

We now describe the theoretical protocol for performing a braiding transformation on a collection of Majorana zero modes exclusively through a sequence of projective measurements, without needing to move the zero modes.  We subsequently describe a teleportation-based measurement protocol for realizing our proposal.  Consider the schematic setup shown in Fig. \ref{fig:Schematic}; Majorana zero modes $\gamma_{1}$, $\gamma_{2}$, $\gamma_{3}$, and $\gamma_{4}$ are used to encode two logical qubits, while $\eta_{1}$ and $\eta_{2}$ will serve, for our purposes, as a single ancilla qubit.  We prepare the ancilla qubit in the state $i\eta_{1}\eta_{2} = +1$ so that the initial state of the system is given by
\begin{align}
\ket{\psi_{i}} = \ket{\phi}\otimes\ket{i\eta_{1}\eta_{2} = +1}
\end{align}
with $\ket{\phi}$, the logical two-qubit state of the four Majoranas $\{\gamma_{i}\}$ that we wish to manipulate.

Our measurement-based braiding protocol is based on the fact that projective measurements of Majorana bilinear operators
\begin{align}
\hat{P}^{(\pm)}_{\gamma_{n}\eta_{m}} \equiv \frac{1 \pm i\gamma_{n}\eta_{m}}{2},
\end{align}
 may be used to implement a unitary braiding transformation up to an overall normalization factor.  Specifically, observe the mathematical identity
\begin{align}
\hat{P}^{(+)}_{\eta_{1}\eta_{2}}\hat{P}^{(+)}_{\eta_{1}\gamma_{3}}\hat{P}^{(+)}_{\gamma_{2}\eta_{1}}\ket{\psi_{i}} = \frac{1}{2^{3/2}}\hat{U}_{23}\ket{\psi_{i}}
\end{align}
where the operator
\begin{align}
\hat{U}_{23} \equiv \frac{1 + \gamma_{2}\gamma_{3}}{\sqrt{2}}
\end{align}
implements the unitary braiding transformation (\ref{rule}).   The measurements that must be performed to realize this braiding operation, starting from the state $\ket{\psi_{i}}$, are summarized in Fig. \ref{fig:Schematic}.

\begin{figure}[t]
\includegraphics[trim = 0 0 0 0, clip = true, width=0.46\textwidth, angle = 0.]{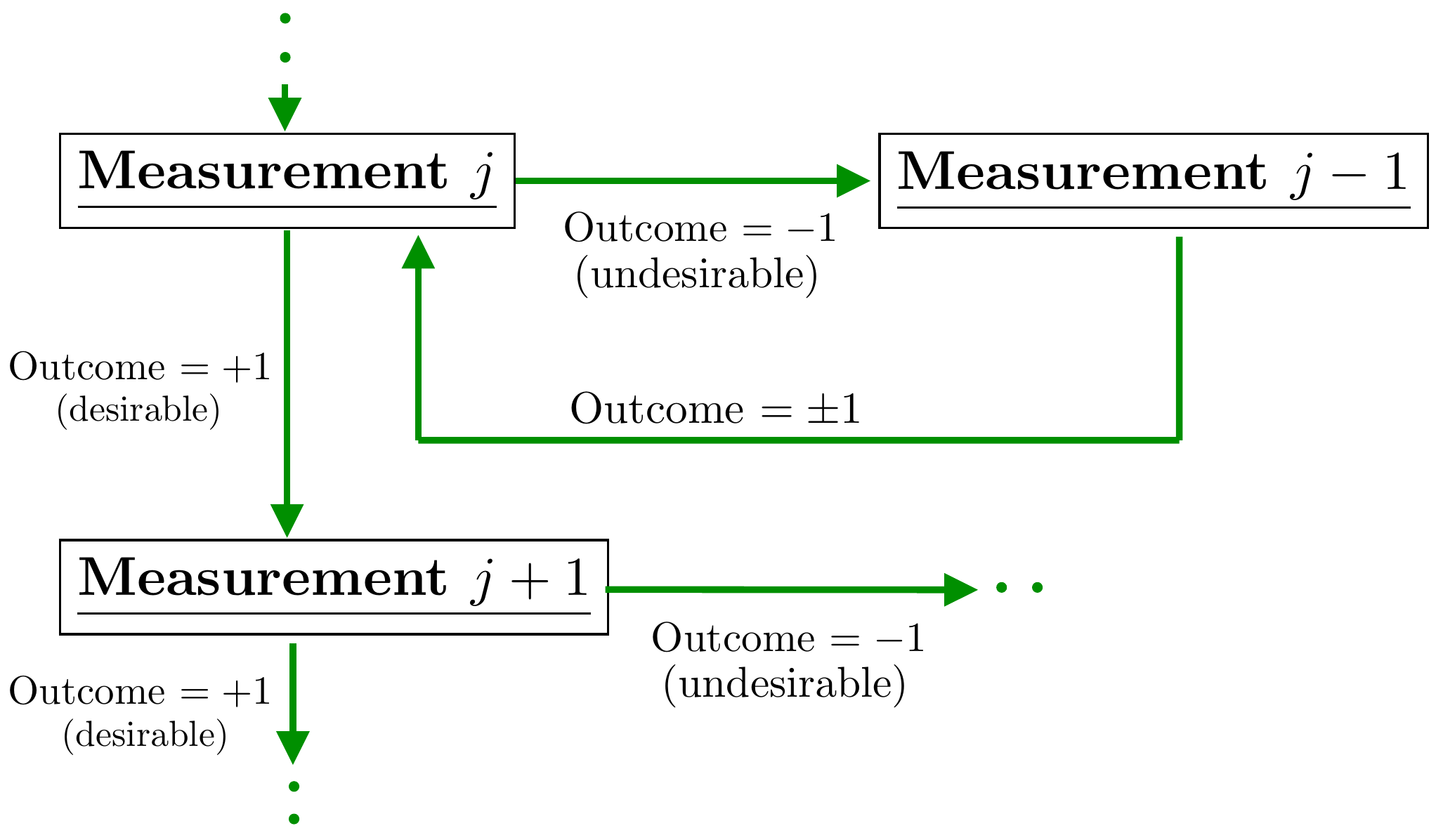}
\caption{{\bf Measurement ``Decision Tree" --} Summary of the measurement protocol.  If a measurement yields an undesirable outcome, the previous measurement step may be repeated (as indicated) to recover the state, before the undesirable measurement was performed. }
  \label{fig:Decision_Tree}
\end{figure}

Successfully performing a measurement-based braiding transformation crucially relies on the {outcomes} of the measurements that are performed.  If a measurement yields an undesirable outcome, however, it is still possible to obtain the desired final state  by performing an appropriate sequence of operations.  
As an example, assume that the first measurement yields the undesirable result that $i\gamma_{2}\eta_{1} = -1$ so that subsequently, the state of the system is given by $\ket{\varphi} \equiv P^{(-)}_{\gamma_{2}\eta_{1}}\ket{\psi_{i}}$.  We may recover the state of the system \emph{before} the undesirable measurement, $\ket{\psi_{i}}$, by measuring the bilinear $i\eta_{1}\eta_{2}$.  If we find that $i\eta_{1}\eta_{2} = +1$, then we recover the initial state
\begin{align}
P^{(+)}_{\eta_{1}\eta_{2}} \ket{\varphi} = \frac{1}{2}\ket{\psi_{i}}
\end{align}
up to a change in normalization, and we may now re-do the measurement of the bilinear $i\gamma_{2}\eta_{1}$.  More generally, in order to recover the state $\ket{\psi_{i}}$, we must alternate measurements of the bilinears $i\eta_{1}\eta_{2}$ and $i\gamma_{2}\eta_{1}$ until we obtain the measurement outcome $i\eta_{1}\eta_{2} = +1$. Observe that
\begin{align}\label{eq:Repeat_Measurements}
P^{(+)}_{\eta_{1}\eta_{2}}P^{(s_{n})}_{\gamma_{2}\eta_{1}}P^{(-)}_{\eta_{1}\eta_{2}}\cdots P^{(s_{1})}_{\gamma_{2}\eta_{1}}P^{(-)}_{\eta_{1}\eta_{2}}\ket{\varphi} = -\frac{s_{n}}{2^{n}}\ket{\psi_{i}}
\end{align}
where $s_{i} = \pm 1$.

A similar protocol may be used to recover from any undesirable measurement outcome.  As summarized in the ``decision tree" in Fig. \ref{fig:Decision_Tree}, when measurement step $j$ is undesirable, we alternate between measurement steps $j-1$ and $j$; this cycle is repeated until measurement step $j$ yields the desired outcome.  The number of steps required to recover from an undesirable measurement only changes the normalization of the final state, as can be seen from Eq. (\ref{eq:Repeat_Measurements}).

We have assumed in our analysis that undesirable measurements only arise due to the inherently probabilistic nature of the measurement-based braiding protocol.  However, an undetected error event (e.g. quasiparticle poisoning) that occurs during the measurement procedure can also lead to an undesirable final state.  If the measurements are performed sufficiently rapidly relative to the poisoning time, these errors may be significantly suppressed, due to the quantum Zeno effect \cite{Zeno}.
A concrete scheme for incorporating error correction into measurement-based braiding will the subject of a forthcoming work \cite{Forthcoming}.

\section{Experimental Realization}

 \begin{figure}
 $\begin{array}{c}
 \includegraphics[trim = 0 0 0 0, clip = true, width=0.47\textwidth, angle = 0.]{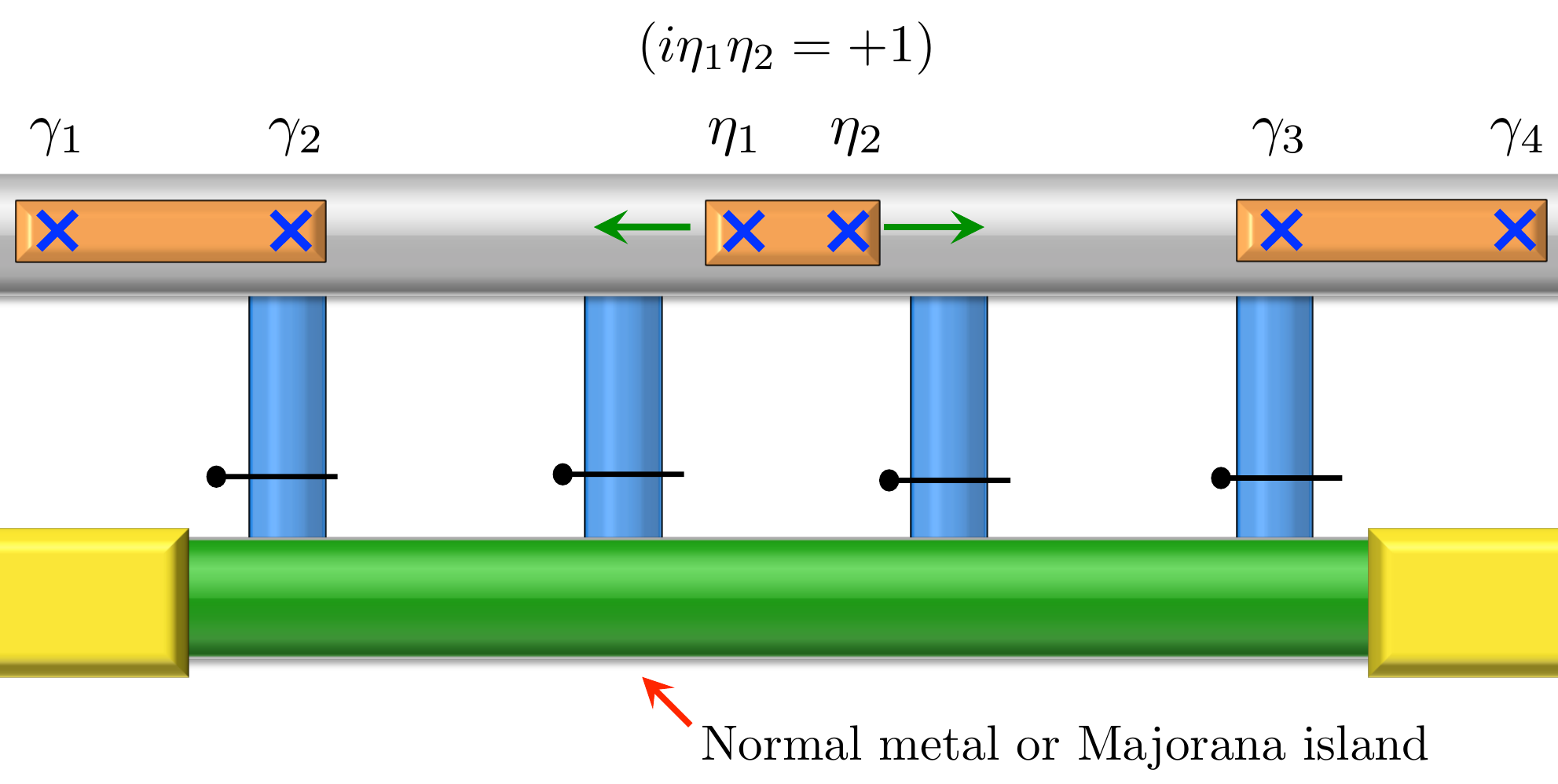}\\\\
\text{\bf (a) {Initialization}}\\\\
\includegraphics[trim = 0 0 0 0, clip = true, width=0.47\textwidth, angle = 0.]{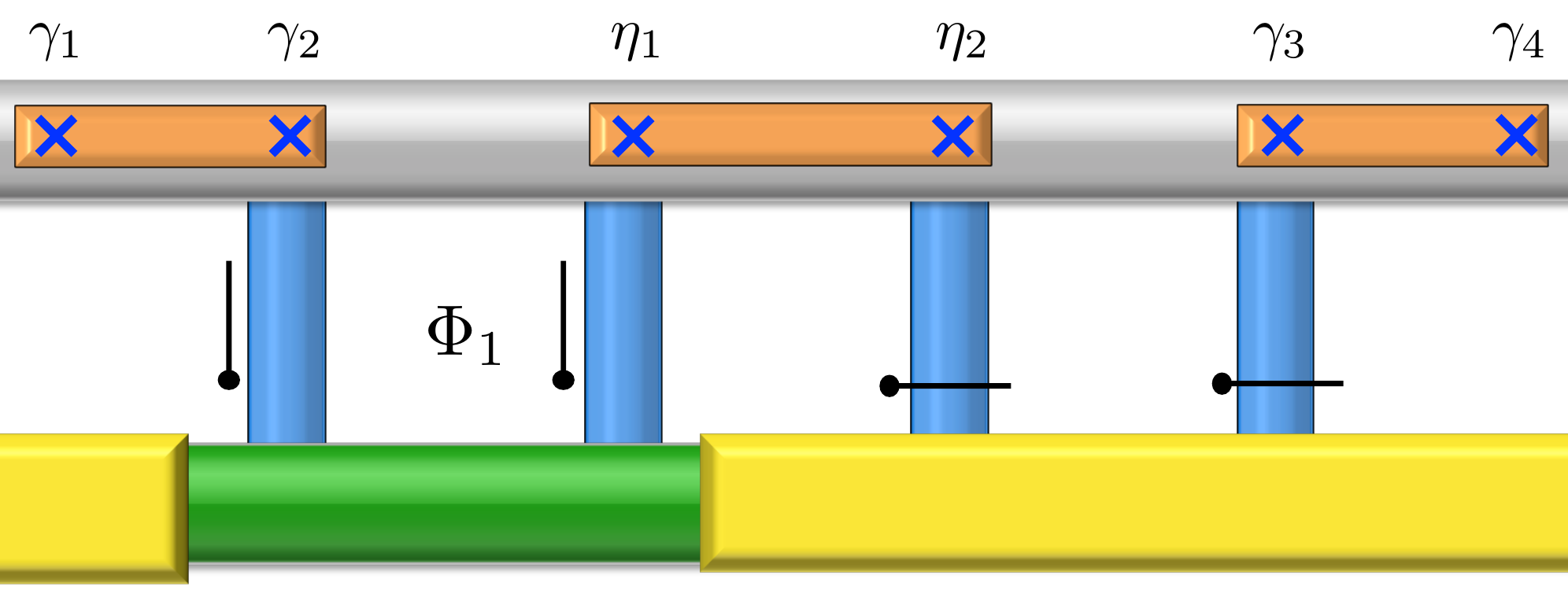}\\\\
\text{\bf (b) {Measurement  I}}\,\,\,(i\gamma_{2}\eta_{1} = +1)\\\\
\includegraphics[trim = 0 0 0 0, clip = true, width=0.47\textwidth, angle = 0.]{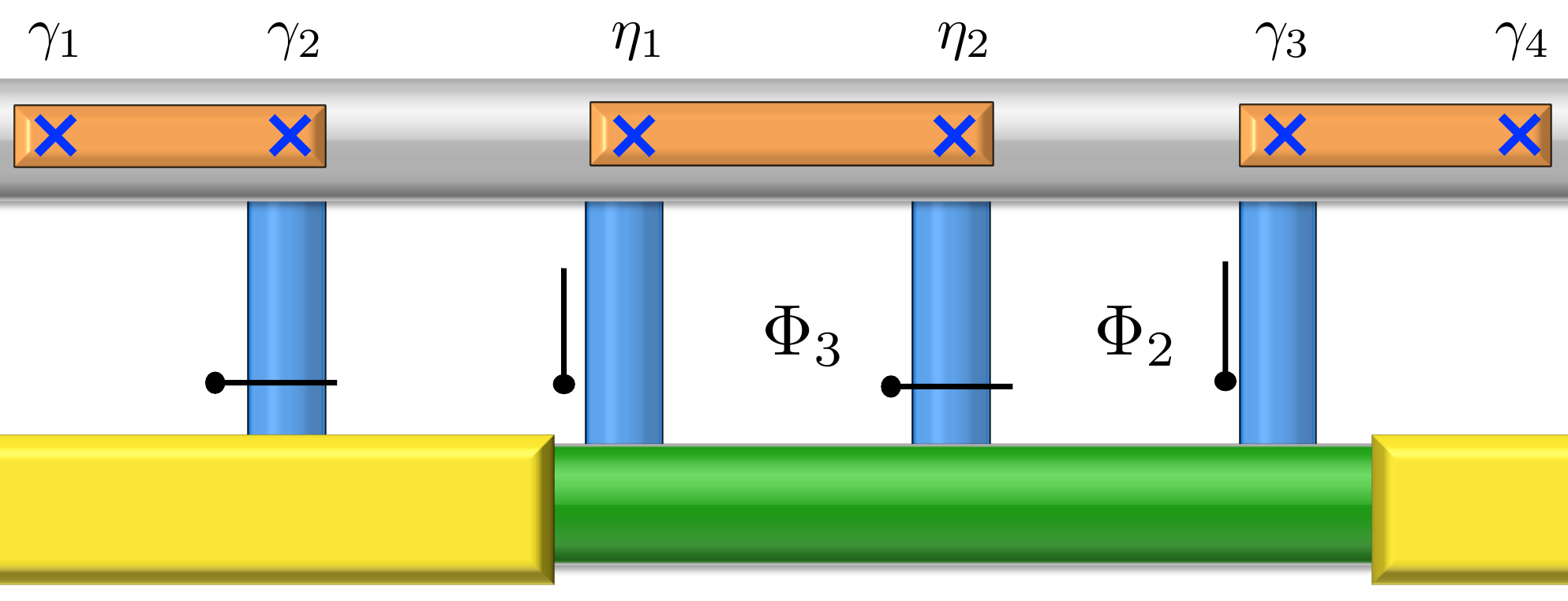}\\\\
\text{\bf (c) {Measurement  II}}\,\,\,(i\eta_{1}\gamma_{3} = +1)\\\\
\includegraphics[trim = 0 0 0 0, clip = true, width=0.47\textwidth, angle = 0.]{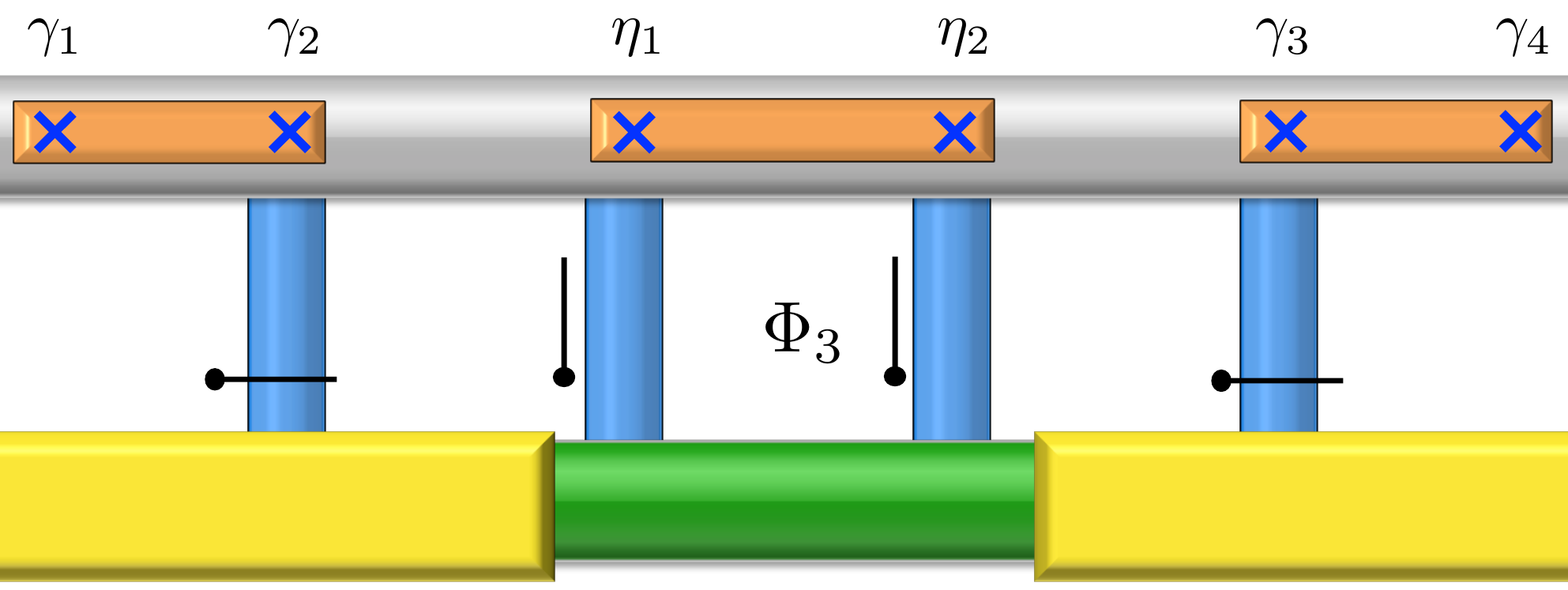}\\\\
\text{\bf (d) {Measurement  III}}\,\,\,(i\eta_{1}\eta_{2} = +1)
\end{array}$
\caption{{\bf Experimental Realization---} Protocol for teleportation-based braiding without braiding is illustrated in a nanowire-based Majorana platform.  A nanowire hosts six Majorana zero modes at the interface between topological and trivial superconducting regions. $\gamma_1,...,\gamma_4$ are used as topological qubits and $\eta_1, \eta_2$ as an ancilla qubit.
The green strip can be either a normal metal with a long phase coherence length or a Majorana island in a definite parity state (a Majorana bus).  We begin by initializing the the ancilla qubit in (a), before performing measurements of the appropriate Majorana bilinears in the top nanowire.  The coupling between the topological superconductor wire to the normal metal or Majorana bus through the metallic strips may be turned on and off, as indicated schematically by the ``switches''. Fluxes may be applied through appropriate loops for topological qubit readout via conductance or persistent current measurement. }
  \label{fig:Device}
\end{figure}

\subsection{Experimental Setup}
We now propose a teleportation-based scheme for realizing  measurement-based braiding; our proposal is summarized in Fig. \ref{fig:Device}.  Consider a superconducting nanowire; gate voltages may be applied along the length of the nanowire to introduce an interface between the topological and trivial superconducting regions, which localizes a Majorana zero mode.  In our setup, we apply gate voltages so that six Majorana zero modes appear ($\gamma_{1}, \gamma_{2}, \gamma_{3}, \gamma_{4}$, $\eta_{1}$ and $\eta_{2}$) at points along the wire.  The distance between the Majoranas is assumed to be sufficiently large, so that hybridization between adjacent Majoranas may be neglected.  We initialize $\eta_{1}$ and $\eta_{2}$ in the state $i\eta_{1}\eta_{2} = +1$ by nucleating the two Majorana zero modes in a topologically trivial region of the nanowire where the total fermion parity is fixed.

Parallel to the existing nanowire in our setup, we now place either (i) a normal metal strip or (ii) a single, proximitized nanowire with gate voltages applied appropriately so that the gated region is topological and hosts two Majorana zero modes ($\psi_{1}$ and $\psi_{2}$) at its two ends.  In both setups, four metal bridges are used to connect the existing nanowire to the normal metal or the second nanowire. In the following section, we will describe the implementation of setup (ii).  A similar protocol may be used to implement setup (i), involving the same sequence of interferometric or flux-based measurements, as long as the metal strip has a sufficiently long phase coherence length.

To implement our ``braiding without braiding" protocol using setup (ii), we will tune gate voltages to  re-position $\psi_{1}$ and $\psi_{2}$ along the length of the second nanowire; hence we will refer to this nanowire as the ``Majorana bus".   We initialize the Majorana zero modes in the bus in the state $i\psi_{1}\psi_{2} = +1$.   The Majorana bus and the remaining Majoranas in our setup are coupled together by four metallic bridges, and each coupling can be tuned on or off, as indicated schematically by the ``switches" in Fig. \ref{fig:Device}.  Each lead is chosen to be shorter than the phase coherence length of the metal, to allow for Majorana qubit readout based on electron teleportation and intereference.

We may perform projective measurements of Majorana bilinears in the top nanowire using the interference of electron trajectories through our setup. This can be achieved by measuring the persistent current in a closed loop, or by measuring the two-terminal conductance across the Majorana bus.
To implement either measurement procedure, we first align the Majorana bus so that $\psi_{1}$ and $\psi_{2}$ are across from the pair Majorana zero modes that we wish to measure, respectively. By turning on the switches on the metallic strips, we introduce electron tunneling between aligned Majorana zero modes on the bus and on the top wire.  In the presence of a large charging energy on both the bus and the wire, the effective Hamiltonian describing the full system is given by Eq.(\ref{Hct}); it depends periodically on the flux through the loop $\Phi$ and on the eigenvalue of the Majorana bilinear operator in the top wire. A flux- or conductance-based readout of the Majorana bilinear may then be performed, as detailed in the following section.


\subsection{Measurement Procedures}
To perform a flux-based measurement one of the Majorana bilinears, we tune gate voltages in the Majorana bus so that $\psi_{1}$ and $\psi_{2}$ are across from the Majorana zero modes in the top row that we wish to measure.  For concreteness, consider a measurement of $i\gamma_{2}\eta_{1}$, as shown in Fig. \ref{fig:Device}a.  After aligning the Majorana bus, we turn off the couplings in the last two metallic strips; this is indicated schematically by the closed and open switches in Fig. \ref{fig:Device}a.  Furthermore, we insert a flux $\Phi_{1}$ through the loop formed by $\psi_{1}$,
$\psi_{2}$, $\eta_{1}$ and $\gamma_{2}$, as shown.

 In the presence of a charging energy that removes the degeneracy between even and odd charge-states on both the Majorana bus and on the top nanowires, the Hamiltonian for the system is $H = H_{t} + H_{c}$ -- with
 \begin{align}
 H_{t} =   it_{1} \psi_{1}\gamma_{2} \,e^{i (\theta_1 - \theta_2)/2}  + i t_{2} e^{i e\Phi/\hbar} \psi_{2}\eta_{1}\, e^{i (\theta_2 -\theta_1)/2} + \mathrm{h.c.}\nonumber
 \end{align} describing the coupling between the Majorana bus to $\gamma_{2}$ and $\eta_{1}$ through the metallic strips, while $H_{c}$ is the charging energy on the nanowire and Majorana bus, as given previously in Eq. (\ref{eq:H_c}). A measurement of the persistent current may be used to determine the Majorana bilinear $i\gamma_{2}\eta_{1}$ as detailed in Sec. IIA.

To perform a conductance measurement, we may introduce a weak tunnel coupling between the Majorana bus and two external leads. A similar protocol for measuring stabilizer operators for the Majorana fermion surface code \cite{Egger, Maj_Surf_Code, Maj_Surf_Code_2, Egger_Roadmap} has also been proposed \cite{Egger, Egger_Roadmap}. The tunneling Hamiltonian takes the form
\begin{align}
H_{T} = t_{L}c^{\dagger}_{L}\psi_{1}e^{-i\theta_{1}/2} + t_{R}c^{\dagger}_{R}\psi_{2}e^{-i\theta_{1}/2}  + \mathrm{h.c.}
\end{align}
where $t_{L,R}$ are the tunnel couplings to the left and right leads.  Here,  $e^{\pm i\theta_{1}/2}$ is the charge-$e$ raising (lowering) operator on the bus, while $c^{\dagger}_{L}$ and $c^{\dagger}_{R}$ are the electron creation operators in the left and right lead, respectively.


When the charging energy is large, we may derive an effective Hamiltonian that takes the form $H_{\mathrm{eff}} = H_{0} + H_{1}$ to lowest order, where
\begin{align}
H_{0} = \frac{2|t_{1}t_{2}|}{E_{+} + E_{-}}\cos\left[\frac{e(\Phi_{1} - \Phi_{c})}{\hbar}\right]\psi_{1}\psi_{2}\gamma_{2}\eta_{1}
\end{align}
and
\begin{widetext}
\begin{align}
&H_{1} = t_{L}^{*}t_{R}\left(\frac{1}{E_{+}} + \frac{1}{E_{-}}\right)\psi_{1}\psi_{2}c^{\dagger}_{R}c_{L} + 2|t_{1}t_{2}|t_{L}^{*}t_{R}\left[\frac{1}{(E_{+})^{3}} + \frac{1}{(E_{-})^{3}}\right]\cos\left[ \frac{e(\Phi - \Phi_{c})}{\hbar}\right](i\gamma_{2}\eta_{1})c^{\dagger}_{R}c_{L} + \mathrm{h.c.}\nonumber
\end{align}
\end{widetext}
The tunneling conductance depends sensitively on the measured value of the bilinear $i\gamma_{2}\eta_{1}$ and is determined to be
\begin{align}
G = g_{0} + \frac{2\pi e^{2}}{\hbar}g_{1}\cos\left[\frac{e(\Phi_{1} - \Phi_{c})}{\hbar}\right]\psi_{1}\psi_{2}\gamma_{2}\eta_{1}.
\end{align}
Here $g_{0}$ is a constant contribution to the conductance that is independent of the measurement outcome, while
\begin{align}
g_{1} = {4|t_{L}|^{2}|t_{R}|^{2}|t_{1}t_{2}|}\left[\frac{1}{(E_{+})^{3}} + \frac{1}{(E_{-})^{3}}\right]\left[\frac{1}{E_{+}} + \frac{1}{E_{-}}\right] \rho_{L}\rho_{R}\nonumber
\end{align}
with $\rho_{L,R}$, the density of states in the left and right leads, respectively.\\


\section*{Acknowlegements}
We thank Anton Ahkmerov and Charlie Marcus for interesting discussions. This work is supported by DOE Office of Basic Energy Sciences, Division of Materials Sciences and Engineering under Award DE-SC0010526. LF is supported partly by the David and Lucile Packard Foundation.  SV is supported partly by the KITP Graduate Fellows Program.

\appendix
\section{Supplemental Material}

To estimate the magnitude of the persistent current in a Majorana interferometer with a metallic arm or another Majorana island, we compute the persistent current in a free electron ring with one and two weak links.


\subsection{Single Weak Link}

Consider a free electron ring of length $L$ with a single weak link.  The bosonized form of the action $S = S_{0} + S_{\mathrm{weak}}$ where $S_{0}$ is \cite{Haldane_1, Haldane_2}
\begin{align}
S_{0} = \frac{1}{2v_{F}}\int_{0}^{L}dx\int_{0}^{\beta}d\tau \left[ (\partial_{\tau}\phi)^{2} + v_{F}^{2}(\partial_{x}\phi)^{2}\right]
\end{align}
with $v_{F}$ is the Fermi velocity of the metal. The ``weak link" is modeled by a weak hopping between the ends of the ring \cite{Kane_Fisher_1, Kane_Fisher_2} $S_{\mathrm{weak}} = -\int d\tau\, \left[\widetilde{t}\,\psi^{\dagger}(L,\tau)\psi(0, \tau) + \mathrm{h.c.}\right]$, with $\psi$ and $\psi^{\dagger}$ the electron creation/annihilation operators respectively.  After bosonizing, the most relevant term in the action for the weak hopping given by
\begin{align}
S_{\mathrm{weak}} = -\widetilde{t}\int_{0}^{\beta}d\tau\,\cos\left[\sqrt{\pi}\,(\phi(L,\tau) - \phi(0,\tau)) + \Theta\right]\nonumber
\end{align}
with
\begin{align}
\Theta \equiv \frac{2\pi\Phi}{\Phi_{0}}
\end{align}
the flux through the ring, in units of the flux quantum $\Phi_{0} = h/e$.

We observe that the fields $\phi(x,\tau)$ for $x\ne 0$, $L$ may be integrated out to obtain an effective action that only involves the phase difference $\vartheta(\tau) \equiv \frac{1}{2}\left[\phi(L,\tau) - \phi(0,\tau)\right]$ \cite{Kane_Fisher_1, Kane_Fisher_2}.  Expanding the phase difference as
\begin{align}
\vartheta(\tau) = \frac{1}{\beta}\sum_{i\omega_{n}}e^{i\omega_{n}\tau}\vartheta(\omega_{n})
\end{align}
with Matsubara frequencies $\omega_{n}\equiv 2\pi n/\beta$, and integrating out the bulk fields results in the effective action for a finite-size system
\begin{align}
S_{\mathrm{eff}} = &\frac{1}{\beta}\sum_{i\omega_{n}}|\omega_{n}|\coth\left[\frac{|\omega_{n}|L}{2v_{F}}\right]|\vartheta(\omega_{n})|^{2}\nonumber\\
& - \widetilde{t}\int_{0}^{\beta}d\tau\,\cos\left[2\sqrt{\pi}\,\,\vartheta(\tau) + \Theta\right]
\end{align}
We may determine the persistent current $I = \beta^{-1}(\partial\ln Z/\partial\Phi)$ perturbatively in the coupling $\widetilde{t}$, where $Z = \int D\phi \,e^{-S_{\mathrm{eff}}}$ is the path integral.  The leading contribution to the persistent current comes at $O(\widetilde{t})$; observe that
\begin{align}
I = \frac{1}{\beta}\frac{\partial\ln Z}{\partial \Phi} &= \frac{e\widetilde{t}}{\beta}\int_{0}^{\beta}d\tau\,\left\langle \sin\left[2\sqrt{\pi}\,\,\vartheta(\tau) + \Theta\right]\right\rangle_{0} + \cdots\nonumber \\
&= e\widetilde{t} \,\sin\left(2\pi\frac{\Phi}{\Phi_{0}}\right)e^{-2\pi\left\langle \vartheta(\tau)^{2}\right\rangle} + \cdots
\end{align}
where $\langle\cdots\rangle_{0}$ denotes the expectation value with respect to the Gaussian part of the action.  From the two-point correlation function
\begin{align}
\langle \vartheta(\tau)\vartheta(0) \rangle_{0} = \frac{1}{\beta}\sum_{i\omega_{n}}\frac{e^{i\omega_{n}\tau}}{ {|\omega_{n}|\coth\left[\frac{|\omega_{n}|L}{2v_{F}}\right]}}
\end{align}
we determine that the persistent current at \emph{zero temperature} is given, to leading order in the weak hopping, and after restoring factors of $\hbar$, by
\begin{align}
I = -e\widetilde{t}\left(\frac{\pi\hbar v_{F}}{\epsilon_{F}L}\right)e^{-\gamma}\sin\left(2\pi \frac{\Phi}{\Phi_{0}}\right)+ O(t^{2})
\end{align}
where $\gamma$ is the Euler-Mascheroni constant. Here, $\hbar \widetilde{t}$ -- the hopping strength across the Majorana island -- is given by $\hbar \widetilde{t} = t^{2}(1/E_{+} + 1/E_{-})$ with $E_{\pm}$ the energy difference between adjacent charge states on the Majorana island as defined in the main text, and $t$, the tunneling amplitude into a single Majorana zero mode.  We may re-express the persistent current in terms of $E_{\pm}$ as well as the tunnel-coupling $\Gamma$ and the level-spacing in the metallic wire $\delta$.  The persistent current is then given by $I = I_{0}\sin(2\pi\Phi/\Phi_{0})$ with
 \begin{align}
{ I_{0} = \frac{ 2e\Gamma\delta}{\hbar} \left(\frac{1}{E_{+}} + \frac{1}{E_{-}}\right)e^{-\gamma}}
 \end{align}
 Taking the tunnel-coupling $\Gamma$ to be approximately one-tenth of the superconducting gap $\Delta$, which determines the level-broadening of the Majorana edge-states 
 we find that $I_{0} \approx 10$ nA.
\subsection{Double Weak Link}
A similar answer is obtained in the case of a metallic wire with two weak links, which we model as two independent wires, each of length $L/2$, whose ends are weakly coupled. In  this case, the bosonized action is given by $S = S_{0} + S_{\mathrm{weak}}$ with
\begin{align}
S_{0} = \frac{1}{2v_{F}}\int_{0}^{L/2}dx\int_{0}^{\beta}d\tau \sum_{\ell=1,2}\left[ (\partial_{\tau}\phi_{\ell})^{2} + v_{F}^{2}(\partial_{x}\phi_{\ell})^{2}\right]
\end{align}
and
\begin{align}
S_{\mathrm{weak}} = &-\widetilde{t}_{1}\int_{0}^{\beta}d\tau\,\cos\left[\sqrt{\pi}\,(\phi_{2}(0,\tau) - \phi_{1}(L/2,\tau)) + \Theta\right]\nonumber\\
&-\widetilde{t}_{2}\int_{0}^{\beta}d\tau\,\cos\left[\sqrt{\pi}\,(\phi_{1}(0,\tau) - \phi_{2}(L/2,\tau))\right]
\end{align}
with $\Theta$ again given by the flux through the ring $\Theta = 2\pi\Phi/\Phi_{0}$.

We now integrate out the bulk fields after expanding the phase differences $\vartheta_{1}(\tau) \equiv [\phi_{2}(0,\tau) - \phi_{1}(L/2,\tau)]/2$ and $\vartheta_{2}(\tau) \equiv [\phi_{1}(0,\tau) - \phi_{2}(L/2,\tau)]/2$ in terms of Matsubara frequencies
$\vartheta_{\ell}(\tau) = {\beta}^{-1}\sum_{i\omega_{n}}e^{i\omega_{n}\tau}\vartheta_{\ell}(\omega_{n})$ to obtain the effective action $S = S_{0}^{(\mathrm{eff})} + S_{\mathrm{weak}}^{(\mathrm{eff})}$ where
\begin{widetext}
\begin{align}
&S_{0}^{(\mathrm{eff})} = \frac{1}{\beta}\sum_{i\omega_{n}}|\omega_{n}|\coth\left(\frac{|\omega_{n}|L}{2v_{F}}\right)\left\{ |\vartheta_{1}(\omega_{n})|^{2} + |\vartheta_{2}(\omega_{n})|^{2} + \mathrm{sech}\left(\frac{|\omega_{n}|L}{2v_{F}}\right)\left[ \vartheta_{1}^{*}(\omega_{n})\vartheta_{2}(\omega_{n}) + \mathrm{c.c.}\right]\right\}\\
&S_{\mathrm{weak}}^{(\mathrm{eff})} = -\widetilde{t}_{1}\int_{0}^{\beta}d\tau\,\cos\left[2\sqrt{\pi}\,\vartheta_{1}(\tau) + \Theta\right]-\widetilde{t}_{2}\int_{0}^{\beta}d\tau\,\cos\left[2\sqrt{\pi}\,\vartheta_{2}(\tau)\right]
\end{align}
\end{widetext}
We again compute the persistent current perturbatively in the hoppings $\widetilde{t}_{1,2}$.  We begin by observing that in terms of the variable
\begin{align}
\varphi(\omega_{n}) = \vartheta_{1}(\omega_{n}) + \mathrm{sech}\left(\frac{|\omega_{n}|L}{2v_{F}}\right)\vartheta_{2}(\omega_{n})
\end{align}
the Gaussian part of the action is now diagonal
\begin{align}
S_{0}^{(\mathrm{eff})} = \frac{1}{\beta}\sum_{\omega_{n}}|\omega_{n}| &\Big[ \coth\left(\frac{|\omega_{n}|L}{2v_{F}}\right)|\varphi(\omega_{n})|^{2} \nonumber\\
&+ \tanh\left(\frac{|\omega_{n}|L}{2v_{F}}\right)|\vartheta_{2}(\omega_{n})|^{2}\Big].
\end{align}
The two-point correlation function
\begin{align}\label{eq:Two_Pt_1}
\left\langle\vartheta_{2}(\tau)^{2}\right\rangle_{0} = \frac{1}{\beta}\sum_{\omega_{n}} |\omega_{n}|^{-1} \coth\left(\frac{|\omega_{n}|L}{2v_{F}}\right)
\end{align}
diverges at low frequencies, so that
\begin{align}
\left\langle\cos\left[2\sqrt{\pi}\,\vartheta_{2}(\tau)\right]\right\rangle_{0} = 0.
\end{align}
As a result, the leading contribution to the persistent current comes at $O(\widetilde{t}_{1}\widetilde{t}_{2})$, as expected.

To leading order in the weak hopping, the persistent current now takes the form
\begin{align}
I = \frac{e\widetilde{t}_{1}\widetilde{t}_{2}}{\beta}\int_{0}^{\beta}d\tau d\tau'\left\langle \sin\left[2\sqrt{\pi}\,\vartheta_{1}(\tau) + \Theta\right]\cos\left[2\sqrt{\pi}\,\vartheta_{2}(\tau')\right]\right\rangle_{0}\nonumber
\end{align}
We may explicitly evaluate the above the expression, after using the correlation function (\ref{eq:Two_Pt_1}) and the fact that
\begin{align}
\left\langle\vartheta_{1}(\tau)\vartheta_{2}(0)\right\rangle_{0} = -\frac{1}{\beta}\sum_{\omega_{n}} \frac{e^{i\omega_{n}\tau}}{|\omega_{n}|\sinh\left(\frac{|\omega_{n}|L}{2v_{F}}\right)}.
\end{align}
At zero temperature, however, the perturbative expression for the persistent current does not converge, as the small parameter in the perturbative expansion turns out to be $\widetilde{t}_{1}\widetilde{t}_{2}\beta^{2}$.  Instead, we find that at low temperatures $T$, the persistent current takes the form $I = I_{0}\sin(2\pi\Phi/\Phi_{0})$ where $I_{0}$, after re-writing in terms of the level-spacing $\delta$, the charging-energy cost for the islands $E_{\pm}$, and the tunnel-coupling for each Majorana island $\Gamma_{i}$ is given by
\begin{align}
{I_{0} = \frac{8\pi e}{\hbar}\frac{\Gamma_{1}\Gamma_{2}}{k_{B}T}\delta^{2}\left(\frac{1}{E_{+}} + \frac{1}{E_{-}}\right)^{2}e^{-2\gamma}}.
\end{align}


\begin{thebibliography}{1}




\bibitem{Kitaev}
A. Kitaev, Phys. Usp. {\bf 44}, 131 (2001).

\bibitem{moore-read}
G. Moore and N. Read, Nucl. Phys. B {\bf 360}, 362 (1991).

\bibitem{read-green}
N. Read and D. Green, Phys. Rev. B {\bf 61}, 10267 (2000).

\bibitem{ivanov}
D. A. Ivanov, Phys. Rev. Lett. {\bf 86}, 268 (2001).

\bibitem{Nayak-rmp}
C. Nayak, S. H. Simon, A. Stern, M. Freedman, and S. Das Sarma, Rev. Mod. Phys. {\bf 80}, 1083 (2008).

\bibitem{Beenakker} C. W. J. Beenakker, Annu. Rev. Con. Mat. Phys. {\bf 4}, 113 (2013).


\bibitem{Alicea}
J. Alicea, Rep. Prog. Phys. {\bf 75}, 076501 (2012).




\bibitem{Delft} V.  Mourik,   K.  Zuo,   S.  M.  Frolov,   S.  R.  Plissard, E. P. A. M. Bakkers,  and L. P. Kouwenhoven, 
    Science {\bf 336},  1003-1007 (2012).

\bibitem{Princeton} S.  Nadj-Perge,  I.  K.  Drozdov,  J.  Li,  H.  Chen,  S.  Jeon,
J. Seo, A. H. MacDonald, B. A. Bernevig, and A. Yazdani, 
Science {\bf 346}, 602 (2014).

\bibitem{Vortex_Tunneling} Hao-Hua Sun, Kai-Wen Zhang, Lun-Hui Hu, Chuang Li, Guan-Yong Wang, Hai-Yang Ma, Zhu-An Xu, Chun-Lei Gao, Dan-Dan Guan, Yao-Yi Li, Canhua Liu, Dong Qian, Yi Zhou, Liang Fu, Shao-Chun Li, Fu-Chun Zhang and Jin-Feng Jia, 
    Phys. Rev. Lett. {\bf 116}, 257003 (2016).

\bibitem{Fu_Kane}
L. Fu and C. L. Kane, Phys. Rev. Lett. {\bf 100}, 096407 (2008).

\bibitem{T_Junction} J. Alicea, Y. Oreg, G. Refael, F. von Oppen, and M. P. A. Fisher, Nature Phys. {\bf 7}, 412 (2011).


\bibitem{Milestone_Majorana}  D. Aasen, M. Hell, R. V. Mishmash, A. Higginbotham,
J. Danon, M. Leijnse, T. S. Jespersen, J. A. Folk, C. M. Marcus, K. Flensberg, and J. Alicea, 
arXiv:1511.05153.

\bibitem{DiVincenzo} F. L. Pedrocchi and D. P. DiVincenzo, 
Phys. Rev. Lett. {\bf 115}, 120402 (2015).

\bibitem{DiVincenzo_2} F. L. Pedrocchi, N. E. Bonesteel and D. P. DiVincenzo,  	Phys. Rev. B {\bf 92}, 115441 (2015).

\bibitem{Sau}  J. D. Sau, D. J. Clarke, and S. Tewari, Phys. Rev. B {\bf 84}, 094505 (2011).

\bibitem{Akhmerov}  B. van Heck, A. R. Akhmerov, F. Hassler, M.Burrello, and C. W. J. Beenakker,  New J. Phys. {\bf 14}, 035019 (2012).

\bibitem{Oppen} Torsten Karzig, Falko Pientka, Gil Refael, Felix von Oppen, Phys. Rev. B {\bf 91}, 201102(R) (2015)




\bibitem{Teleportation} L. Fu, 
    Phys. Rev. Lett {\bf 104}, 054602 (2009).



\bibitem{Nielsen_Chuang} M. A. Nielsen and I. L. Chuang, Phys. Rev. Lett. {\bf 79}, 321 (1997).

\bibitem{Nielsen} M. A. Nielsen, 
Physics Letters A {\bf 308}, 2 (2003).

\bibitem{Bonderson} P. Bonderson, M. Freedman, and C. Nayak, 
Phys. Rev. Lett. {\bf 101}, 010501 (2008).


\bibitem{Copenhagen_2}  S.  M.  Albrecht,  A.  P.  Higginbotham,  M.  Madsen,  F.
Kuemmeth,  T.  S.  Jespersen,  J.  Nygard,  P.  Krogstrup, and C. M. Marcus,
Nature {\bf 531}, 206 (2016).



\bibitem{Nanowire_Proposal}  J. D. Sau, R. M. Lutchyn, S. Tewari, and S. das Sarma,
Phys. Rev. Lett. {\bf 104}, 040502 (2010).

\bibitem{oreg}
Yuval Oreg, Gil Refael, and Felix von Oppen, Phys. Rev. Lett. {\bf 105}, 177002 (2010)

\bibitem{palee}
A. C. Potter and P. A. Lee, Phys. Rev. Lett. {\bf 105}, 227003 (2010).

\bibitem{Egger-PRL}
R. Hutzen, A. Zazunov, B. Braunecker, A. L. Yeyati and R. Egger, Phys. Rev. Lett. {\bf 109}, 166403 (2012).

\bibitem{Glazman_1}  B. van Heck, R.M. Lutchyn, L.I. Glazman, 
    Phys. Rev. B {\bf 93}, 235431 (2016).

\bibitem{Marcus-private} C. M. Marcus, private communication (2016).

\bibitem{Sondhi} V. Khemani, R. Nandkishore and S. Sondhi, Nat. Phys. {\bf 11}, 560-565 (2015).

\bibitem{Rahul} S. Johri and R. Nandkisore, arXiv:1608.00022v1 preprint.

\bibitem{Copenhagen} A. P. Higginbotham, S. M. Albrecht, G. Kirsanskas, W. Chang, F. Kuemmeth, P. Krogstrup, T. S. Jespersen, J. Nygard, K. Flensberg, C. M. Marcus, 




\bibitem{Xu-Fu} C. Xu and L. Fu, Phys. Rev. B {\bf 81}, 134435 (2010)


\bibitem{Egger} L. A. Landau, S. Plugge, E. Sela, A. Altland, S. €‰M. Albrecht, and R. Egger
Phys. Rev. Lett. {\bf 116}, 050501  (2016).

\bibitem{Maj_Surf_Code} S. Vijay, T. H. Hsieh and L. Fu, Phys. Rev. X {\bf 5}, 041038 (2015).

\bibitem{Maj_Surf_Code_2} S. Vijay and L. Fu, Phys. Scr. {\bf 2016}, T168 (2016).

\bibitem{Egger_Roadmap}  S. Plugge, L. A. Landau, E. Sela, A. Altland, K. Flensberg, R. Egger, arXiv:1606.08408.

\bibitem{Zeno} B. Misra and E. C. G. Sudarshan, J. Math. Phys. {\bf 18}, 756 (1977).

\bibitem{Teo} J. C. Y. Teo, J. Phys.: Condens. Matter, {\bf 28}, 143001 (2016).

\bibitem{Halperin} B. I. Halperin, Y. Oreg, A. Stern, G. Refael, J. Alicea, and F. von Oppen
Phys. Rev. B {\bf 85}, 144501 (2012).

\bibitem{Forthcoming} S. Vijay and L. Fu, unpublished.

\bibitem{SM} Supplemental Material.

\end{thebibliography}

\begin{thebibliography}{1}

\bibitem{Haldane_1} F. D. M. Haldane, J. Phys. C {\bf 14}, 2585 (1981).

\bibitem{Haldane_2} F. D. M. Haldane, Phys. Rev. Lett. {\bf 47}, 1840 (1981).

\bibitem{Kane_Fisher_1}
C. L. Kane and M. P. A. Fisher, Phys. Rev. Lett. {\bf 68}, 1220 (1992)

\bibitem{Kane_Fisher_2}
C. L. Kane and M. P. A. Fisher, Phys. Rev. B {\bf 46}, 15233 (1992).

\end{thebibliography}
\end{document}